\documentclass[preprint,12pt,authoryear]{elsarticle}

\usepackage{graphicx}
\usepackage{float} 
\usepackage{booktabs} 
\usepackage{subcaption} 
\usepackage{amsmath} 
\usepackage{amssymb}  
\usepackage{amsthm} 
\newtheorem{definition}{Definition}
\newtheorem{proposition}{Proposition}
\newtheorem*{proposition*}{Proposition}
\usepackage{color} 
\usepackage[left]{lineno} 
\usepackage{ulem} 
\usepackage[a4paper, top=1in, bottom=1in, left=0.75in, right=0.75in]{geometry}

\makeatletter
\def\ps@pprintTitle{%
 \let\@oddhead\@empty
 \let\@evenhead\@empty
 \def\@oddfoot{\reset@font\footnotesize\textit{Accepted for publication in Travel Behaviour and Society.}\hfil}%
 \let\@evenfoot\@oddfoot}
\makeatother

\journal{Travel Behaviour and Society}

\begin{document}
\begin{frontmatter}



\title{Understanding everyday public transit travel habits: a measurement framework for the peakedness of departure time distributions}


\author[inst1]{Jiwon Kim}

\affiliation[inst1]{organization={School~of~Civil~Engineering, The~University~of~Queensland},
            addressline={St~Lucia}, 
            city={Brisbane},
            postcode={4072}, 
            state={Queensland},
            country={Australia}}

\author[inst2]{Jonathan Corcoran}

\affiliation[inst2]{organization={School~of~the~Environment, The~University~of~Queensland},
            addressline={St~Lucia}, 
            city={Brisbane},
            postcode={4072}, 
            state={Queensland},
            country={Australia}}

\begin{abstract}
Persuasive scholarship presents how individual daily travel habits implicate congestion, environmental pollution, and the travel experience. However, the empirical characteristics and dynamics of travel habits remain poorly understood. Quantifying both our individual travel habits and how these habits aggregate to form system-wide dynamics are of critical importance to enable smart design of public transit systems that are better tailored to our daily mobility needs. We contribute to this need through the development and implementation of a new measurement framework capturing the `peakedness' of users’ departure time distributions. Departure time `peakedness' reflects a user's tendency to repeatedly choose the same departure time for a given origin-destination trip, offering a clearer and more intuitive representation of regularity and habitual patterns compared to traditional metrics like standard deviation or entropy. Our framework demonstrates that system-wide departure time peakedness can be decomposed into individual users’ departure time peakedness and the alignment of their peak times. This allows for a systematic analysis of both individual and collective behaviours. We apply our framework to departure time data from a 12-month period, encompassing 5,947,907 bus journeys made by 29,640 individuals across three urban networks within a large regional metropolis. Our findings reveal that departure time peakedness is more deeply tied to inherent, passenger-specific characteristics, such as passenger type, rather than external factors like weather or holidays. Additionally, individual-level departure time peakedness shows notable dynamics over time, indicating that habitual routines can evolve in the long term, while system-level peakedness exhibits remarkable long-term stability.
\end{abstract}




\end{frontmatter}


\section{Introduction}
\label{sec:intro}


Human mobility habits are a reflection of how our daily lives are framed by a series of regularised activities. These activities are each located in space and time and act to constrain how, where, when and why we move around. When considered in aggregate across a transport system exert important implications for congestion \citep{mannering1994temporal}, environmental pollution \citep{schwanen2012rethinking} and the travel experience \citep{domarchi2008effect}.

Scholarship points to the way that daily travel patterns are repetitive, and individuals tend to recurrently follow the same travel choices \citep{garling2003introduction}. We know that individual trip making is conditioned by a range of factors such as gender \citep{gordon1989gender}, age \citep{chudyk2015destinations}, occupation \citep{rasouli2015employment}, household characteristics \citep{dieleman2002urban}, socio-economics \citep{kotval2015socio} and is further governed by characteristics of the built environment \citep{hong2014built}. However much of this work tends to be cross sectional in nature and does not attempt to capture the longer-term histories of individual travellers that is an essential ingredient to understand travel habits \citep{garling2003introduction}. Thus, if we are to first capture and then empirically measure the way in which travel habits impact our transit systems from which smart solutions to invoke ‘good’ and ‘sustainable’ travel habits can be implemented then we must begin to assemble a longitudinal dimension to travel behaviour scholarship.

Smart card systems employed to automate fare collection have emerged as valuable source of information on travel behaviour with important strategic, tactical, and operational benefits \citep{pelletier2011smart}. Studies have exploited these data to investigate travel demand \citep{morency2007measuring}, trip purpose \citep{lee2014trip}, geographic patterns \citep{tao2014exploring}, temporal travel patterns \citep{cats2022unravelling}, community structure \citep{yildirimoglu2018identification}, service reliability \citep{ma2015modeling}, modal transfer behaviour \citep{sun_characterizing_2015}, and the passenger experience \citep{chu2016reproducing}, to name a select few. Critically, as it relates to understating travel habits, these data offer a continuous record of transit usage through which we can move beyond the traditional cross-sectional studies and use of travel diaries to assemble longitudinal histories of individual travellers.

In this study we draw on a large smart card database of bus ridership and extend the work on travel behaviour by examining long term travel trajectories for individual passengers across a metropolis. More specifically, we develop a novel measurement framework to unveil travel habits by examining the departure time choice behaviours of individual bus riders over a 12-month period. In the transport literature, travel habits has been largely studied in the context of travel mode choice \citep{bamberg2003choice, gardner2009modelling, cherchi2011accounting, sharmeen2014walking} and route choice \citep{bogers2005joint, kurauchi2014variability, he2014daytoday, kim2017route}, but research in the context of departure time choice \citep{thorhauge2020habit} is rare. We aim to fill this gap by developing a new approach to measure habitual aspects of day-to-day departure time distribution in terms of bus riders’ tendency towards choosing the same departure time repeatedly—that we term `peakedness'. This `peakedness' shares conceptual similarities with the `stickiness' of route choice behaviour defined by \citet{kim2017route}, which captures the tendency of bus riders to consistently choose or `stick to' the same route. While the `stickiness index' introduced by \citet{kim2017route} could theoretically be applied to departure time choice, it is primarily designed for `discrete' route choice alternatives, motivating us to devise a new measure that directly addresses the `continuous' nature of departure time choice.

\begin{figure}[htbp]
    \centering
    \includegraphics[width=0.8\textwidth]{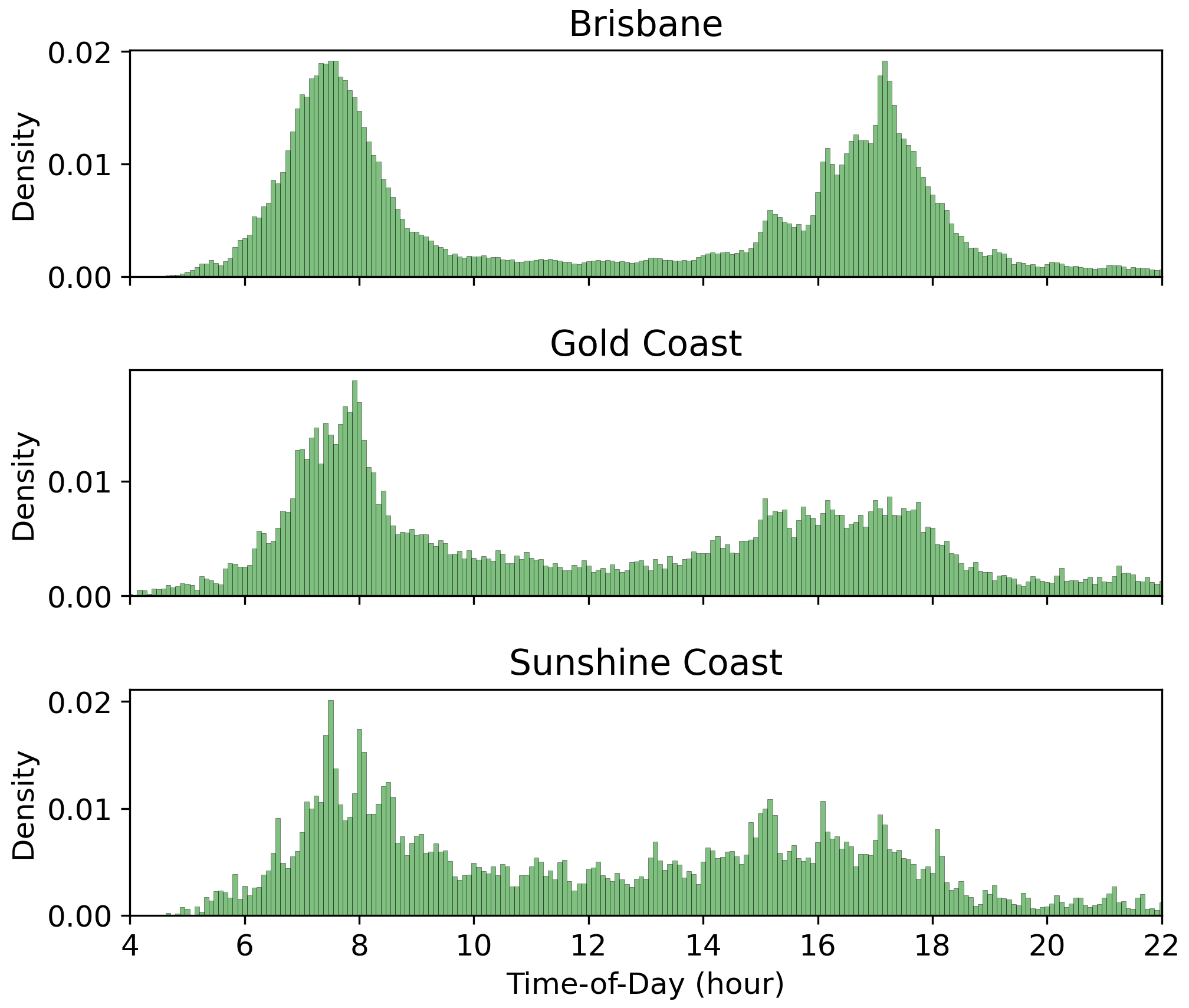} \caption{Bus passenger trips in three major cities in Queensland, Australia, namely Brisbane, Gold Coast, and Sunshine Coast: city-wide departure time distributions for bus commuters are constructed from a 12-month (July 2015 to June 2016) record of \textit{go} card, an electronic smart card ticketing system.}
    \label{fig:departure_histo_sys}
\end{figure}

Departure time distribution, which shows the number of trip departures across different times of the day, provides valuable insights into when and how many trips are initiated. This information is crucial for network operators, transport planners, and service providers to understand peak demand periods, enabling them to align resources, services, and infrastructure to meet varying travel demands throughout the day. The departure time distribution can be constructed for an individual traveller by combining day-to-day departure time choices over a certain period (e.g., 1 year) or can be constructed at the system-wide by combining departure time distributions from all the travellers in the system, forming the system-wide demand loading profile. Figure \ref{fig:departure_histo_sys} shows examples of city-wide departure time distributions for bus commuters, constructed using a 12-month smart card data from three major cities in Queensland, Australia, namely Brisbane, Gold Coast, and Sunshine Coast. While traditional statistics like \textit{standard deviation} or \textit{kurtosis} can provide insights into the shape of a distribution by measuring the `spread' or `tailedness' characteristics, they do not directly measure the `peakedness' of the distribution. In our study, we introduce indicators specifically designed to capture the `peakedness' characteristics of the departure time distribution. Moreover, we derive an analytical expression that offer an intuitive explanation on how the `peakedness' of individual travellers' departure time distributions affect the `peakedness' of the overall system-wide departure time distribution.
These indicators are simple and easy to apply, yet insightful, aiming to provide a more robust and systematic approach to analysing the degree of concentration of trips during specific times of the day. The concept of `peakedness' and associated measurement framework offer a new and important lens through which to examine travel habits of individual users across a metropolis and how these habits scale to form metropolitan-wide dynamics.

\section{Methodology}

\subsection{A Measure of Departure Time Peakedness}
We define the `peakedness' of a departure time distribution as the extent to which departure time observations are concentrated around the peak period or the most preferred departure time. A highly peaked distribution indicates that a significant proportion of the trips occur within the defined peak window. In other words, the more peaked the distribution, the greater the concentration of trips within this peak time window. Given a time window of $ h $-minutes and a probability density function $ f $ of departure times, we define the \textit{$ h $-minute Peak Trip Concentration} (PTC), denoted by $ \psi_h $, as the maximum probability that the departure time $X$ falls within any $ h $-minute window as follows:
\begin{equation} \label{eq:psi_h}
    \psi_h \equiv \max_t P([t,t+h]) = \max_t \int_{t}^{t+h} f_X(x) \,dx
\end{equation}
where $X$ is a real random variable representing the departure time for a given individual trip, $t$ is a specific time of day---both $X$ and $t$ are expressed in \textit{total minutes since the start of the day}---, and $P([t,t+h])$ represents the probability that departure time $X$ falls within interval $[t,t+h]$, i.e., $P(t \leq X \leq t+h)$.

\begin{figure}[htbp]
    \centering
    \includegraphics[width=0.7\linewidth]{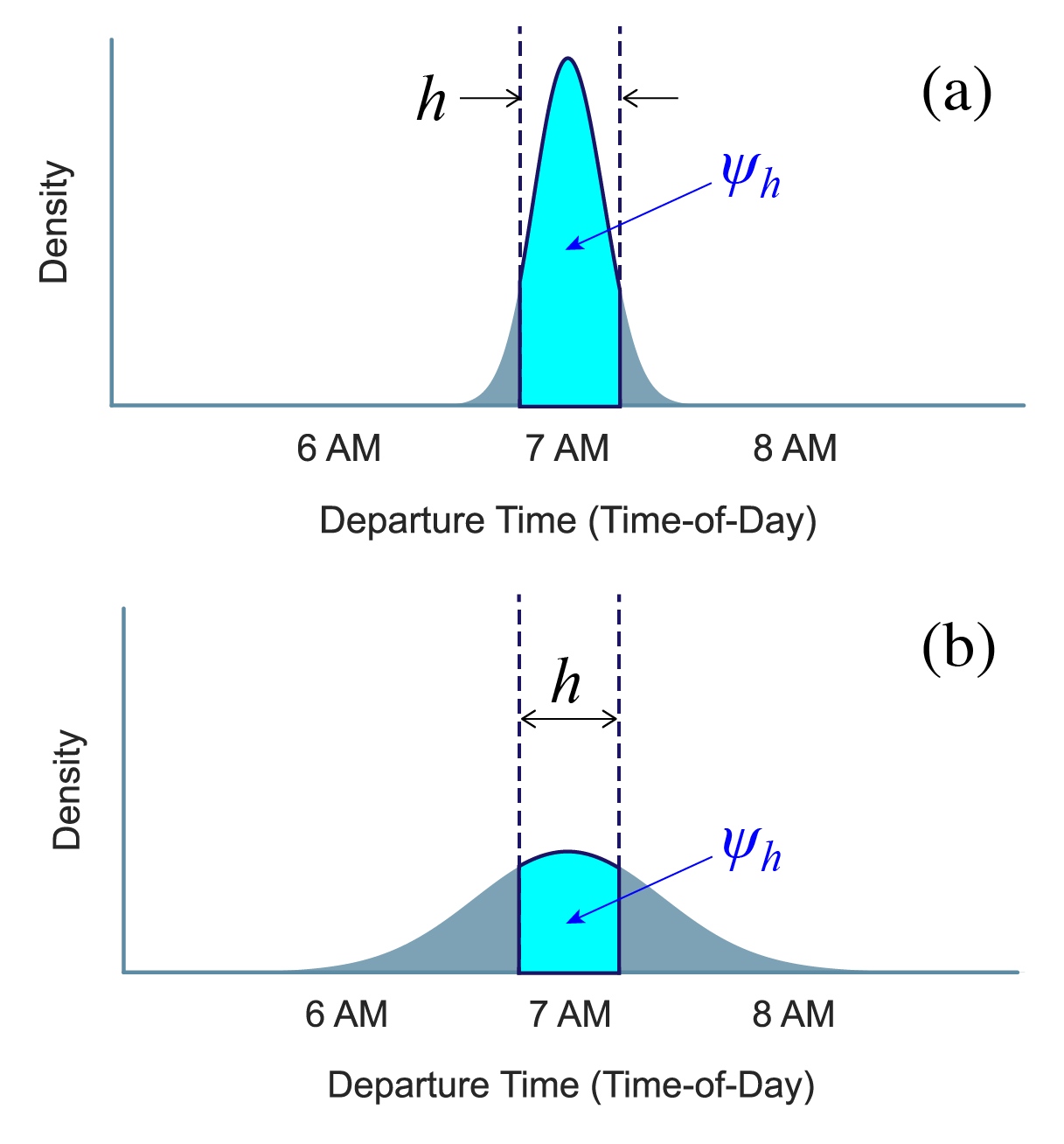}
    \caption{Measuring the $ h $-min Peak Trip Concentration, $ \psi_h $, as an indicator of the \textit{peakedness} of a departure time distribution. The area under the density curve within the $ h $-min peak window represents $ \psi_h $: \textbf{(a)} the value of $ \psi_h $ is larger when the distribution is more peaked and \textbf{(b)} the value of $ \psi_h $ is smaller when the distribution is less peaked. }
    \label{fig:psi_illust}
\end{figure}

Eq.[\ref{eq:psi_h}] finds the time window $[t, t+h]$ that contains the largest proportion of departure times, and the corresponding interval probability $ \psi_h $ represents the proportion of trips that fall within the most peaked $ h $-minute window. Figure \ref{fig:psi_illust} illustrates this concept, where departure time distributions are shown in the period of 6-8AM, peaked around 7AM. Then, the area under the density curve within the window of $ h $-minutes centring at 7AM represents the value of the $ h $-min Peak Trip Concentration, $ \psi_h $. When a departure time distribution is highly peaked, this proportion is larger and, thus, the peakedness of the distribution is characterised as a larger $ \psi_h $ value (Figure \ref{fig:psi_illust}(a)). When a departure time distribution is less peaked, its peakedness level is lower and characterised as a smaller $ \psi_h $ value (Figure \ref{fig:psi_illust}(b)).  The value of $ \psi_h $ ranges between 0 and 1, where $ \psi_h $ is 1 (the largest) when all departure times happen within the $ h $-min peak window and $ \psi_h $ is close to 0---more accurately as low as $ \psi_h=h/1440 $ (e.g., $ \psi_h=15/1440=0.01 $ if $ h $ is set to 15 minutes)---when the departure times are uniformly distributed over the whole day (24 hours = 1440 minutes).

To measure departure time peakedness at the individual level, the $ h $-min Peak Trip Concentration, $ \psi_{h,i} $, is estimated for each user $i$. Specifically, $ \psi_{h,i} $ is calculated for each user for a specific origin-destination (OD) pair to evaluate each traveller's own departure time consistency within their individual day-to-day trip distributions for recurring trips. By analysing departure time peakedness for each unique user-OD combination separately, we effectively control for variations in travel distances across users and ODs, ensuring that our method captures how consistently an individual sticks to their preferred departure time window.

Although the proposed metric fundamentally measures the probability of events occurring within a defined window---something not new in probability theory---, the novelty lies in its application as a peakedness metric for departure time distributions. Traditional measures such as standard deviation, kurtosis, or entropy assess the overall characteristics of a distribution but may not fully capture how sharply or consistently departure times are concentrated within a specific interval (see \ref{si:existing_measures}). By quantifying the probability that departures cluster around a specific peak time, the proposed metric offers a robust measure of behaviour that is less sensitive to outliers. It is well-supported by both theory and practice that measuring the concentration of activity within the most intense time period provides a meaningful way to quantify the peakedness of a distribution. Examples that share conceptual similarities with the proposed metric are \textit{relative peakedness} in probability theory (see \ref{si:relative_peakedness}) and \textit{Peak Hour Factor} (PHF) in traffic engineering (see \ref{si:peak_hour_factor}). Additionally, using a probability-based measure allows us to link individual peakedness metrics to system-wide peakedness. Since the system-wide departure time distribution can be expressed as a mixture of individual distributions, the proposed metric enables the exploration of how individual behavioural patterns shape the overall system's departure time patterns. Therefore, while the concept of measuring probabilities is well established, its application to assess individual peakedness in departure time distributions represents a novel and practical contribution.

In this study, we focus on \textit{departure time distributions} rather than \textit{arrival time distributions}, as departure times more directly reflect individual behavioural preferences and habitual travel choices, whereas arrival times are influenced by additional external factors such as congestion and travel time variability. By capturing the repeated selection of preferred departure time windows within each traveller' day-to-day trip choices, our approach provides a clearer measure of how consistently individuals adhere to their habitual travel schedules. Furthermore, a comparative analysis of departure and arrival time peakedness across users, presented in \ref{si:dep_vs_arr}, shows a strong correlation between the two measurement approaches, demonstrating that despite variations in travel time, the overall peakedness profile remains consistent between departure and arrival time distributions.

\subsection{A Measurement Framework for the Peakedness of Travellers' Departure Time Choices and Their Relations to System-wide Dynamics} Using the $ h $-min Peak Trip Concentration ($ \psi_h $) previously defined, we can effectively assess the peakedness of departure time distributions for specific users or groups of users within a given transport system. Next, we seek to elucidate how the proposed peakedness measure reveals a systematic relationship between the system-wide departure time peakedness and the individual user-level departure time peakedness. To begin, we recognise that the departure time distribution for the entire population of travellers in a given system is constructed by aggregating the departure times of individual travellers within the system. As such, we can represent the system-wide departure time distribution as a mixture distribution of all individual travellers' departure time distributions as follows:
\begin{equation} \label{eq:mixture_dist}
P([t,t+h])=\sum_{i=1}^{n} w_i P_i([t,t+h])
\end{equation}
where $P([t,t+h])$ represents the system-wide distribution encompassing a population of $ n $ users, $P_i([t,t+h])$ represents the user-specific distribution for user $ i $, and $ w_i$ is the weight associated with user $ i $ such that $ w_i \geq 0 $ and $ \sum_{i=1}^{n} w_i = 1 $. Substituting Eq.[\ref{eq:mixture_dist}] to Eq.[\ref{eq:psi_h}], we can also express the peakedness measure of the system-wide departure time distribution, denoted by $ \psi_h^{sys} $, in terms of individual users' departure time distributions as follows:
\begin{equation} \label{eq:psi_sys}
\begin{split}
\psi_h^{sys} &= \max_t P([t, t+h])\\
&= \max_t \sum_{i=1}^{n} w_i P_i([t,t+h])
\end{split}
\end{equation}
Then, we propose to divide Eq.[\ref{eq:psi_sys}] by $\sum_{i=1}^{n} \max_{t} w_i P_i([t, t+h]) $ to express the system-level peakedness measure $ \psi_h^{sys} $ in terms of user-level peakedness measure, denoted by $ \psi_{h,i} $, as follows:
\begin{equation} \label{eq:psi_sys_user}
\begin{split}
\psi_h^{sys} & = \frac {\max_{t} \sum_{i=1}^{n} w_i P_i([t, t+h])} {\sum_{i=1}^{n} \max_{t} w_i P_i([t, t+h])} \times \sum_{i=1}^{n} \max_{t} w_i P_i([t, t+h]) \\
& = \frac {\max_{t} \sum_{i=1}^{n} w_i P_i([t, t+h])} {\sum_{i=1}^{n} \max_{t} w_i P_i([t, t+h])} \times \sum_{i=1}^{n} w_i \max_{t} P_i([t, t+h]) \\
& = \frac {\max_{t} \sum_{i=1}^{n} w_i P_i([t, t+h])} {\sum_{i=1}^{n} \max_{t} w_i P_i([t, t+h])} \times \sum_{i=1}^{n} w_i \psi_{h,i}
\end{split}
\end{equation}
where $ \psi_{h,i}=\max_{t} w_i P_i([t, t+h]) $ is the $h$-min Peak Trip Concentration based on user $ i $'s departure time distribution. In Eq.[\ref{eq:psi_sys_user}], $ \psi_h^{sys} $ and  $ \psi_{h,i} $ are connected via the following multiplicative factor, which we refer to as \textit{$h$-min Peak Coincidence Factor} ($ PCF_h $):
\begin{equation} \label{eq:pcf}
\begin{split}
PCF_h &\equiv \frac {\max_{t} \sum_{i=1}^{n} w_i P_i([t, t+h])} {\sum_{i=1}^{n} \max_{t} w_i P_i([t, t+h])} \\
&=\frac {\text{Peak of aggregated load}}{\text{Sum of individual peak loads}}
\end{split}
\end{equation}
This is the ratio of the $ h $-min peak trip concentration of the mixture distribution to the weighted sum of its components' $ h $-min peak trip concentrations. It quantifies the relationship between \textit{the peak of the aggregated trip load} and \textit{the sum of individual peak trip loads}, indicating how closely the peak departure time windows of individual users align. The value of $ PCF_h $ is in the range of 0 to 1 and signifies the degree to which the preferred departure time windows of individual users ($ h $-min peak windows) coincide across the system. When all users' peak departure time windows coincide and their peak loads occur simultaneously within the same $ h $-min window, $ PCF_h $ equals 1, as the system peak load aligns with the sum of the user peak loads. As $ PCF_h $ deviates further from 1 (i.e., the smaller the value of $ PCF_h $), it indicates greater diversity or dispersion in the peak departure windows of individual users. Recognising that $\sum_{i=1}^{n} w_i \psi_{h,i}$ in Eq.[\ref{eq:psi_sys_user}] represents the average of individual users' departure time peakedness, denoted as $ \overline{\psi_h} $, we can simplify Eq.[\ref{eq:psi_sys_user}] as follows:
\begin{equation} \label{eq:main}
\psi_h^{sys} = PCF_h \times \overline{\psi_h}
\end{equation}
The fact that $ \psi_h^{sys} $ is the product of $ PCF_h $ and $ \overline{\psi_h} $ reveals that the peak trip concentration of the system ($ \psi_h^{sys} $) can be decomposed into two components: one associated with \textit{intra-user} peak trip concentrations ($ \overline{\psi_h} $) and the other associated with \textit{inter-user} peak coincidence ($ PCF_h $). This decomposition highlights that the peakedness of the system-wide departure time distribution is influenced by both the individual users’ departure time peakedness ($ \psi_h $) and the alignment of their peak windows ($ PCF_h $). It is worth noting that all three of these measures fall within the range of 0 to 1 and possess clear and intuitive interpretations: the values of $ \psi_h^{sys} $ and $ \overline{\psi_h} $ directly correspond to the probability or trip proportion, while $ PCF_h $ provides a system-to-individual ratio of peak loads, which is readily interpretable. These three measures of peakedness form a novel measurement framework that enables the assessment of the peakedness characteristics of departure time distributions at both the system and user levels and offers insights into their interrelated dynamics.

\begin{figure} [htbp]
    \centering
    \includegraphics[width=\textwidth]{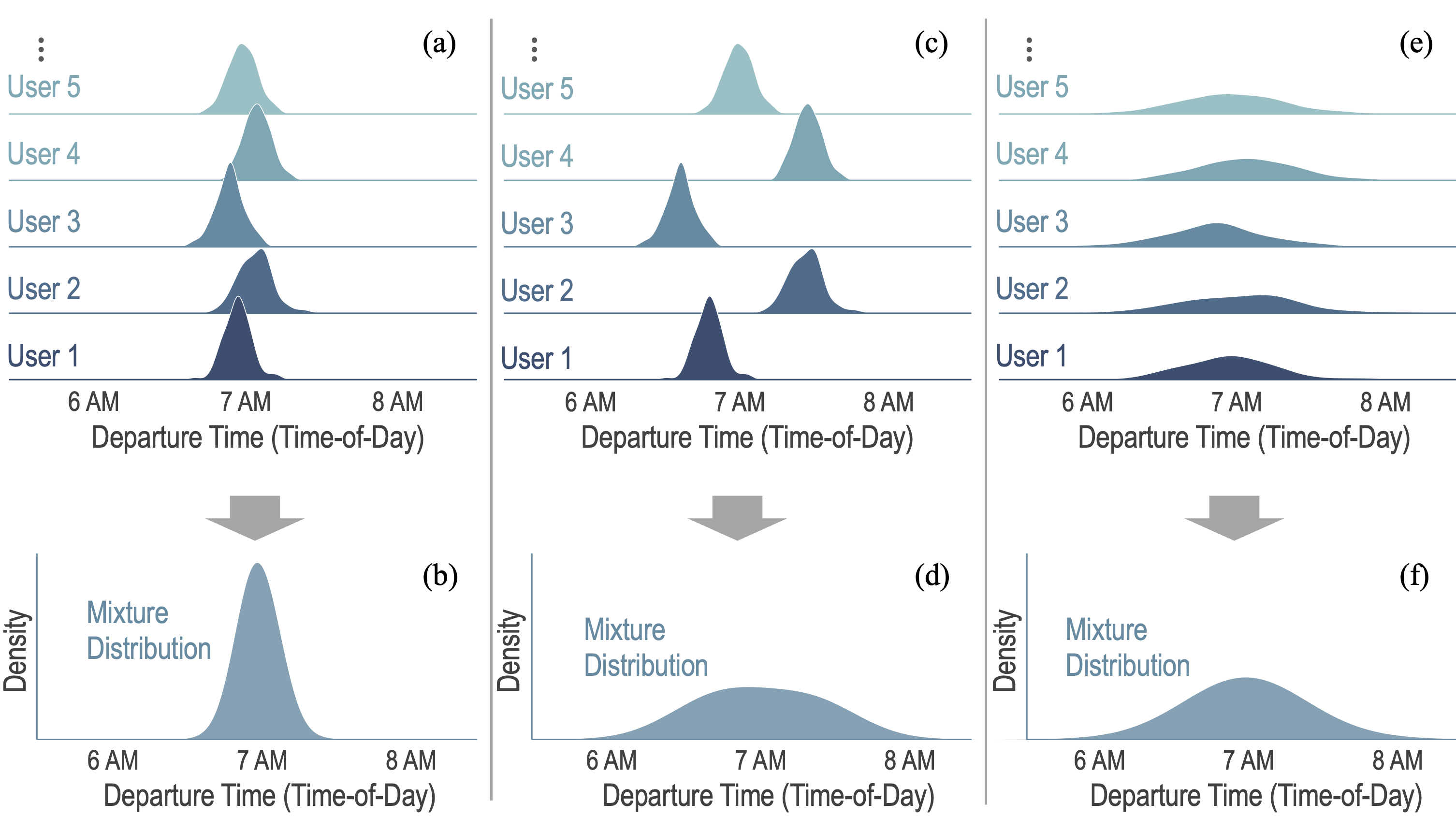}
    \vspace{0.1cm}
    \caption{The relationship among $ \psi_h^{sys}$, $ \overline{\psi_h} $, and $ PCF_h $: [\textbf{a, b}] high user departure time peakedness ($ \overline{\psi_h} $) and high peak alignment ($ PCF_h $) lead to high system-level departure time peakedness ($ \psi_h^{sys}$), [\textbf{c, d}] high user departure time peakedness ($ \overline{\psi_h} $) and low peak alignment ($ PCF_h $) lead to low system-level departure time peakedness ($ \psi_h^{sys}$), [\textbf{e, f}] low user departure time peakedness ($ \overline{\psi_h} $) and high peak alignment ($ PCF_h $) lead to low system-level departure time peakedness ($ \psi_h^{sys}$).}
    \label{fig:mixture_dist}
\end{figure}

Figure \ref{fig:mixture_dist} provides a visual representation of the relationship among these three measures in three different scenarios. In Figure \ref{fig:mixture_dist}(a)-(b), we illustrate a situation where the system-wide departure time distribution exhibits high peakedness (high $ \psi_h^{sys}$) as a result of both the individual users' departure time distributions being highly peaked (high $ \overline{\psi_h} $) and the alignment of their peak periods being high (high $ PCF_h $). Conversely, the other two cases in the figure showcase scenarios where the system-wide departure time has low peakedness. Figure \ref{fig:mixture_dist}(c)-(d) illustrates a situation where this low peakedness arises from a lack of coincidence in the users' peak periods (low $ PCF_h $), despite each user individually having a highly peaked departure time distribution (high $ \overline{\psi_h} $). Figure \ref{fig:mixture_dist}(e)-(f), on the other hand, illustrates a situation where the low system-wide peakedness results from the individual users' departure time distributions having low peakedness (low $ \overline{\psi_h} $), despite their peak periods being highly aligned (high $ PCF_h $).

\subsection{Choice of Peak Window, $h$}\label{sec:choice_of_h}
\begin{figure}[h]
    \begin{subfigure}{0.49\textwidth}
        \includegraphics[width=\textwidth]{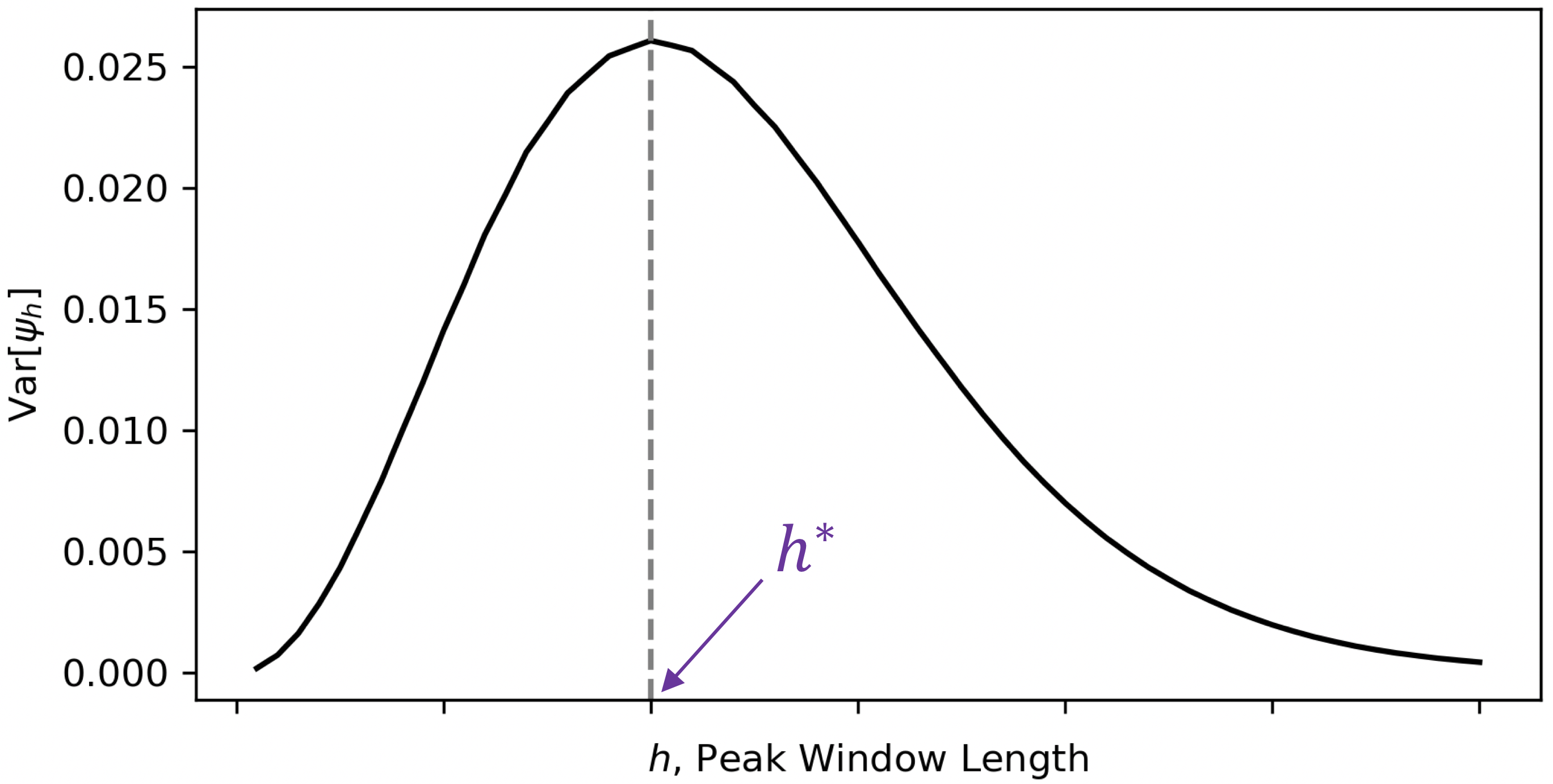}
        \caption{}
        \label{fig:optimal_h_var_max_a}
    \end{subfigure}
    \hfill
    \begin{subfigure}{0.49\textwidth}
        \includegraphics[width=\textwidth]{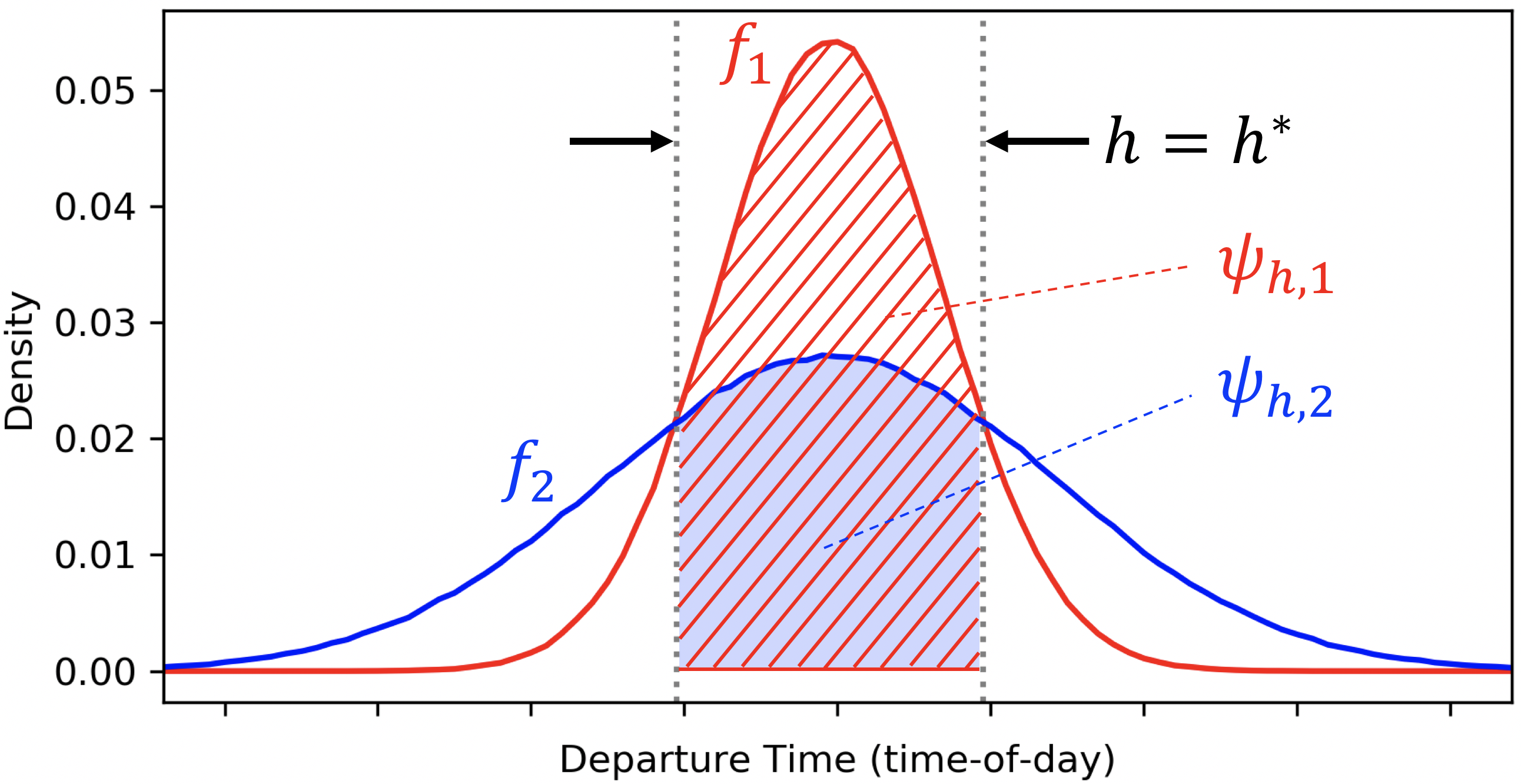}
        \caption{}
        \label{fig:optimal_h_var_max_b}
    \end{subfigure}
\caption{Optimal peak window: (a) the change in the variance of $\psi_h$ across users with respect to different choice of $h$; (b) an illustration of the variance-maximising optimal peak window $ h^* $ in a special case where a system has two peakedness levels, i.e., departure time distribution with high peakedness ($ f_1 $) and departure time distribution with low peakedness ($ f_2 $)}
\label{fig:optimal_h_var_max} 
\end{figure}

Selecting the appropriate width of the peak window, $ h $, is important in measuring and interpreting $ \psi_h $, especially when comparing its values across individual users. If $ h $ is too large, $ \psi_h $ will tend to approach 1, as most departures fall within the wide window, leading to an inflated peakedness for most travellers. Conversely, if $ h $ is too small, $ \psi_h $ will approach 0, since very few departures times fall within the narrow window, resulting in artificially low peakedness for most travellers. A desirable property of $h$ is that the resulting $ \psi_h $ effectively differentiate between different users' varying departure time patterns. As such, we propose to determine the optimal window $ h^* $ as the $ h $ value that maximises the \textit{variance} of $ \psi_h $ ($\text{Var}[\psi_h]$) across users as follows:
\begin{equation} \label{eq:optimal_h}
h^* = \arg\max_h \text{Var}[\psi_h] \mathrel{\hat{=}} \arg\max_h \frac{1}{n-1} \sum_{i=1}^{n} ( \psi_{h,i} - \overline{\psi_h} )^2
\end{equation}
where $ \psi_{h,i} $ is the $h$-min peak trip concentration of user $ i $, $ \overline{\psi_h} $ is
the sample mean of $ \psi_h $ over $ n $ users, and $\text{Var}[\psi_h]$ is the variance of $\psi_h$, which can be estimated ($\mathrel{\hat{=}}$) by calculating the sample variance. Figure \ref{fig:optimal_h_var_max_a} illustrates how the variance of $\psi_h$ changes with respect to different choice of peak window $h$. When $ h $ is very small, the variance is low because $ \psi_h  \approx 0 $ for most users. As $ h $ increases, the variance initially rises but eventually decreases again when $ h $ becomes too large and $ \psi_h  \approx 1 $ for most users. Therefore, the variance of $\psi_h$ as a function of $h$ forms a peak at a critical point, $ h = h^* $, after which it decreases. This critical point, $ h = h^* $, serves as the optimal window width that maximises the spread of  $\psi_h$ values across users, thereby best distinguishing different levels of peakedness in the system. 

In a special case where a system has only two peakedness levels---one group of users with highly peaked departure time distribution $ f_1 $ and another group of users with less peaked departure time distribution $ f_2 $---, the variance-maximising optimal window $ h^* $ corresponds to the peak window at which the two distribution curves intersect, as indicated by the two vertical dotted lines in Figure \ref{fig:optimal_h_var_max_b}. The detailed proof is provided in \ref{si:further_analysis_for_h}. This example offers a useful intuition that the variance-maximising optimal $ h^* $ captures the best peak window that naturally separate different peakedness levels as much as possible, which offers a good middle ground for $h$ that is not too small and not too large given the set of departure time distributions in the dataset.

\section{Results}
\subsection{Data} \label{sed:data}
\begin{figure}[htbp]
    \centering
    \begin{subfigure}{0.5\textwidth}
        \includegraphics[width=\textwidth]{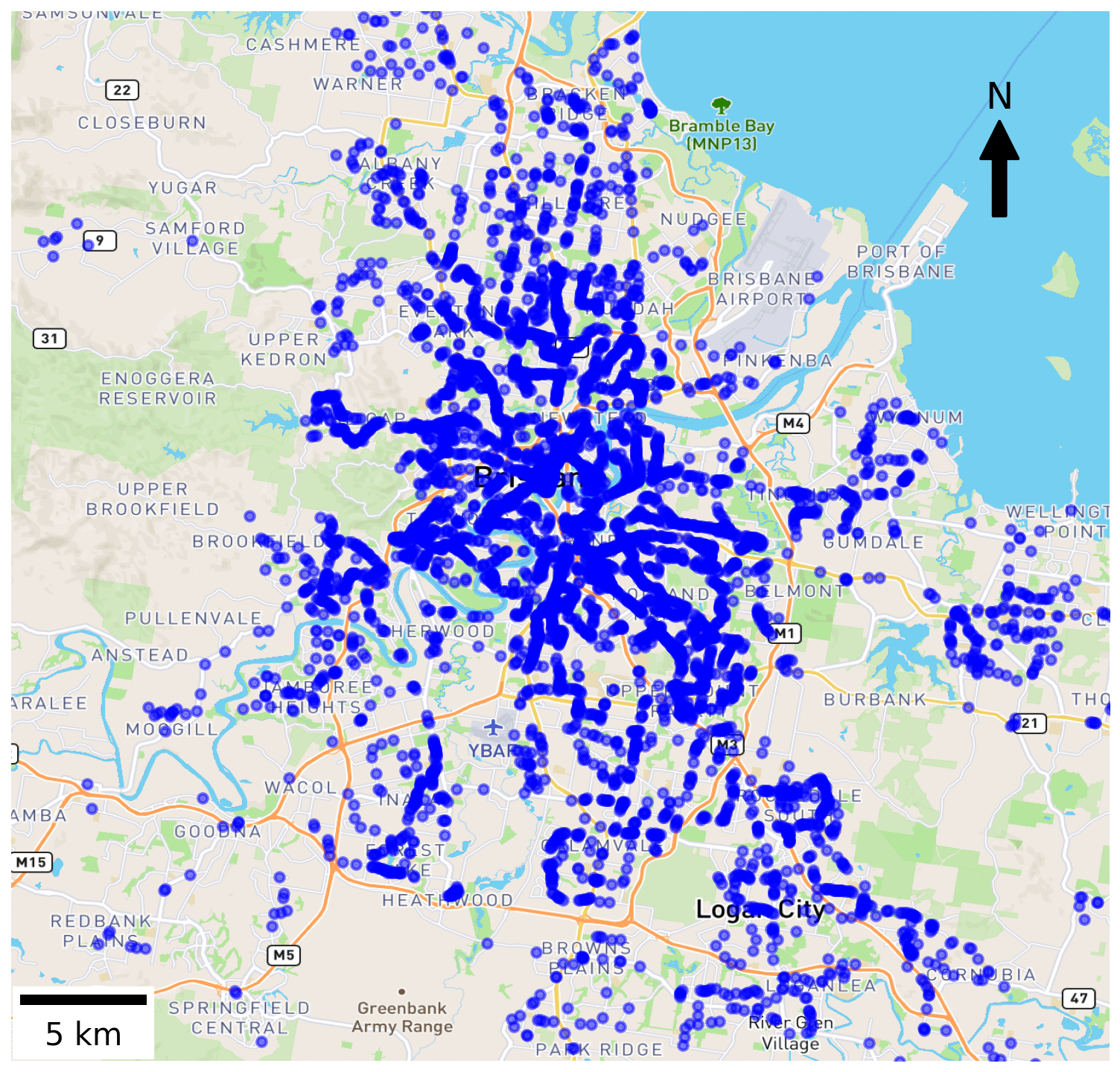}
        \caption{Brisbane}
        \label{fig:map_bne}
    \end{subfigure}
    
    \begin{subfigure}{0.42\textwidth}
        \includegraphics[width=\textwidth]{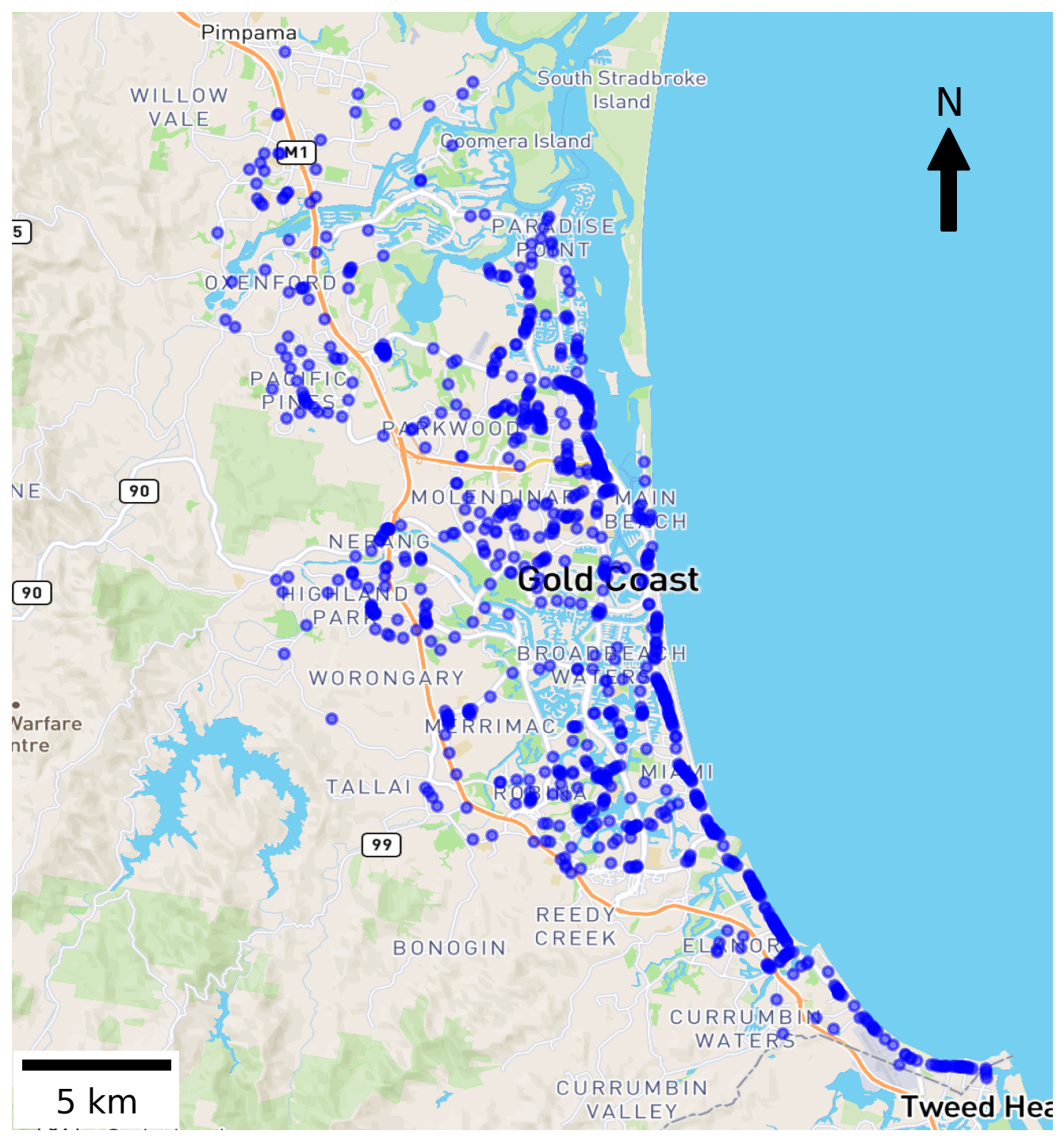}
        \caption{Gold Coast}
        \label{fig:map_gc}
    \end{subfigure}
    \hfill
    \begin{subfigure}{0.53\textwidth}
        \includegraphics[width=\textwidth]{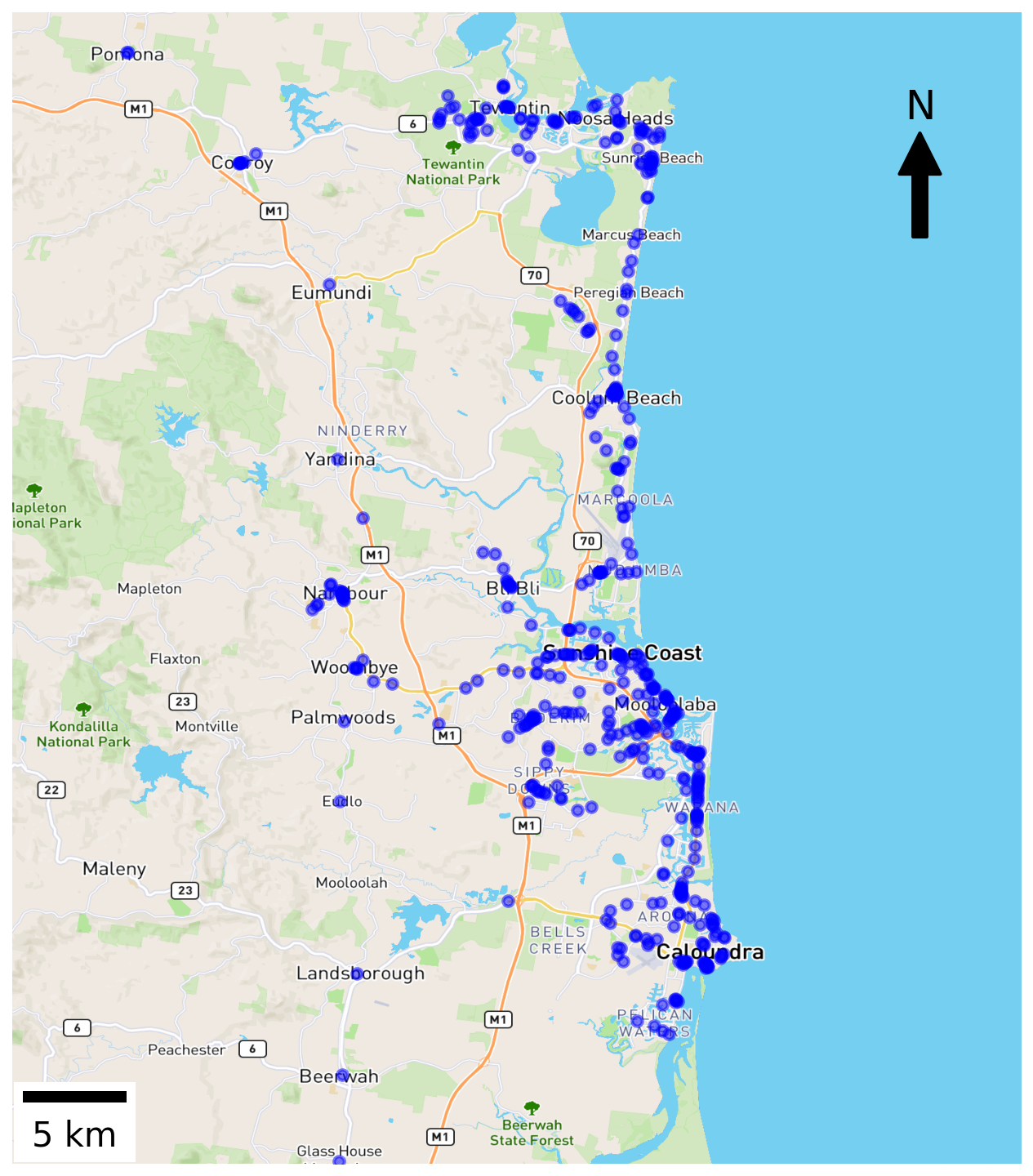}
        \caption{Sunshine Coast}
        \label{fig:map_sc}
    \end{subfigure}
    \caption{Maps of Brisbane, Gold Coast, and Sunshine Coast showing the locations of bus stops (blue dots)}
    \label{fig:maps} 
\end{figure}

The three peakedness measures developed in this work---$ \psi_h $, $ \psi_h^{sys} $, and $ PCF_h $---are estimated using public transport smart card data from South East Queensland (SEQ), Australia, to characterise the departure time peakedness of bus passengers. This study considers three different study networks, namely, Brisbane, Gold Coast, and Sunshine Coast, which are three major cities in the SEQ region, as shown in Figure \ref{fig:maps}.

We use a 12-month (July 2015 to June 2016) record of \textit{go} card, an electronic smart card ticketing system used on the public transport network in SEQ operated by TransLink (the public transit agency for Queensland). The \textit{go} card is the principal means by which passengers pay their transit fares in SEQ, representing approximately 87\% of all trips taken across the TransLink public transport network \citep{translink2016report}. It requires passengers to touch the card on a card reader at the start and end of each \textit{journey} as well as when transferring between \textit{trips}, where TransLink defines a \textit{trip} as the act of travelling from point A to point B with no transfers (a single trip) and a \textit{journey} as the act of travelling from the origin to the final destination that may involve one or a number of trips (transfers) using different transport modes \citep{translink2021gocard}. The \textit{go} card data contain information on each trip's boarding and alighting locations and their associated times, trip ID, journey ID, card ID (\textit{go} card identifier), and passenger type, among other attributes. Passenger types recorded in the \textit{go} card database include \textit{Adult}, \textit{Senior/Pensioner}, \textit{Tertiary Student}, \textit{Secondary Student} (secondary school students aged 15+ years), \textit{Child} (children aged 5 to 14 years), and others (special concession card holders such as veterans).

\begin{figure} [h]
	\centering
	\includegraphics[width=.9\textwidth]{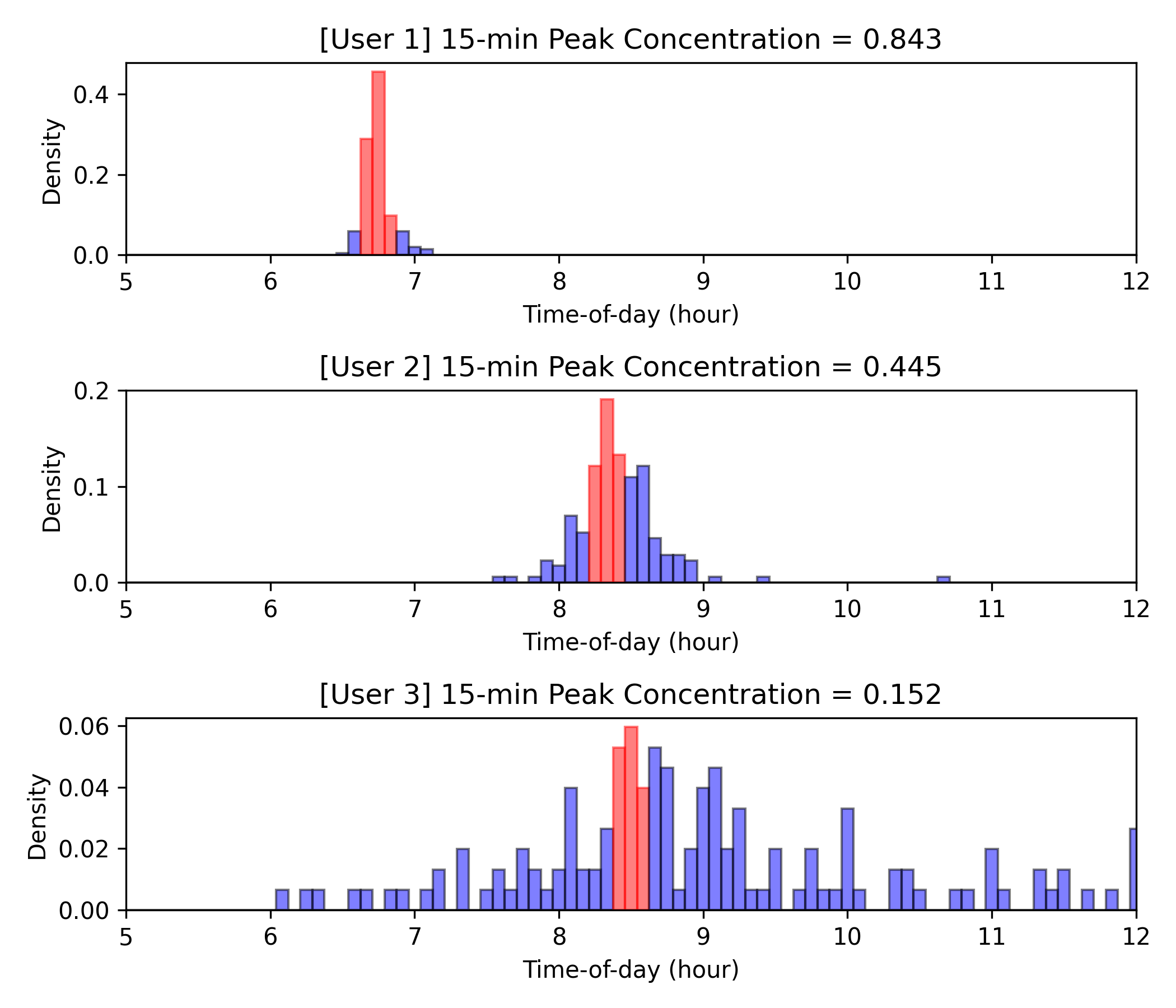}\\
	\caption{Examples of user departure time distributions and the estimated $ h $-min peak trip concentrations, $ \psi_h $, for $ h $=15 minutes.}
	\label{fig:user_hpp_example}
\end{figure}

In this study, we focus on regular bus passengers who use buses repeatedly. For each bus user, trip data are processed and grouped to form journeys for each origin-destination (OD). Then, the departure time distribution is constructed for each user-OD pair by extracting the boarding times at the origin stop for a given OD journey over the course of one year from July 2015 to June 2016. The \textit{user-OD pair} is, thus, the basic unit for measuring the peakedness of user-level departure time distribution, i.e., $ h $-min peak trip concentration for user $i$ ($ \psi_{h,i} $). Each departure time distribution is represented as a normalised histogram with a bin width of 5 minutes, as shown in the examples in Figure \ref{fig:user_hpp_example}. 

Figure \ref{fig:user_hpp_example} shows examples of three bus users' departure time distributions and the estimated $ \psi_h $ for a peak window width of $ h=15 $ minutes. The $ \psi_h $ values for Users 1, 2, and 3 are 0.843, 0.445, and 0.152, respectively. User 1 has the strongest tendency towards choosing a particular departure time as the great majority of his/her trips (84.3\%) occur during the short 15-min peak window. User 2 shows more flexibility in choosing departure time, resulting in about half his/her trips (44.5\%) occurring during the 15-min peak. User 3 has the lowest peakedness and the greatest flexibility among three, showing that his/her most peaked 15-minutes only covers 15.2\% of the trips and the remaining 84.4\% of data are widely spread out from this peak.

To ensure that we have a sufficient sample size to characterise each departure time distribution, we used user-OD pairs that have at least 50 journey data points (i.e., 50 boarding time observations) for constructing a departure time distribution, which roughly captures users making \textit{more than one journey per week} on average for a given OD. When constructing a system-level mixture distribution, $P([t,t+h])=\sum_{i=1}^{n} w_i P_i([t,t+h])$, we combine the departure time distributions of individual user-OD pairs in the system, $P_i([t,t+h])$. We assume all component departure time distributions have equal weights, i.e., $ w_i=\cdots=w_n=1/n $, because we focus on regular users and already filter out insignificant user-OD pairs with too few data points.

Table \ref{table:datasets} shows the summary of dataset size for the three study networks after filtering in terms of the number of users, user-OD pairs, journeys, and origins, with and without grouping by passenger type. The entire data encompasses 5,947,907 bus journeys made by 29,640 individuals across the three study networks. `The number of users' in the table represents the total count of all distinct \textit{go} card users in each dataset. Since each user may have multiple ODs and, thus, result in multiple user-OD pairs, the number of distinct user-OD pairs is greater than the number of users. `The number of journeys' is the sum of the number of journeys across all user-OD pairs. `The number of origins' is the number of unique bus stops from which these journeys started. Overall, the Brisbane dataset is the largest, which has 50,336 user-OD pairs providing 50,336 departure time distributions and their associated ($ \psi_h $) measures, followed by Gold Coast and Sunshine Coast, which has 3968 and 1817 user-OD pairs, respectively. In terms of passenger type, Adult is the largest group in all three cities. The second largest group differs, however, as Tertiary Student is the second largest group in Brisbane and Gold Coast, whereas Senior/Pensioner is the second largest group in Sunshine Coast.

\begin{table}
	\centering
	\caption{Summary of datasets}
	\label{table:datasets} 
	\vspace{-3mm}
	\includegraphics[width=\textwidth]{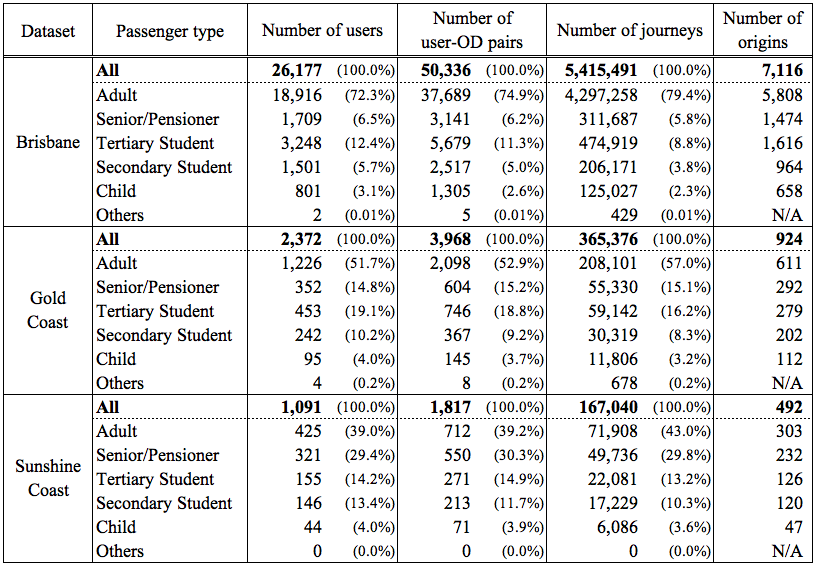}\\
\end{table}

\subsection{Measuring Departure Time Peakedness of Three Cities}\label{sec:results_system} 
In this section, we measure and compare the departure time peakedness of the three study networks. Figure \ref{fig:departure_histo_sys} shows the system-wide departure time distribution for each network, combining all bus commuters' journeys over a one-year period. In Brisbane, two distinct peaks (AM and PM) are observed. Gold Coast and Sunshine Coast display similar dual-peak patterns, although the PM peaks are less sharp compared to Brisbane. Given these two distinct peaks, we separately analyse the peakedness characteristics of the AM and PM periods. 

For each user-OD pair, we first construct trip frequency distributions using 5-minute bins by aggregating trip observations over multiple days (1 year in this study). We then identify the peak 5-minute bin (i.e., the time interval with the highest trip density) for each user-OD pair. The user-OD departure time distributions with their peak 5-minute bin occurring between 12:00 AM and 11:59 AM are classified as AM trips, while those with their peak occurring between 12:00 PM and 11:59 PM are classified as PM trips. For each time-of-day group, we estimate three peakedness measures---$ \overline{\psi_h} $, $ \psi_h^{sys} $, and $ PCF_h $---using the associated user-OD departure time distributions. Individual $h$-min peak trip concentration ($ \psi_{h,i} $) is first computed for each user-OD pair by summing trip densities over an $h$-minute interval centred around the peak 5-minute bin. To obtain the average user-level peakedness ($\overline{\psi_h}$), we then take the mean of $\psi_{h,i}$ across all user-OD pairs within a given network and time-of-day group. Separately, we aggregate all individual user-OD departure time data to construct the system-level mixture distribution and estimate system-level peakedness ($\psi_h^{sys}$) by summing densities over an $h$-minute interval centred around the peak 5-minute bin of this distribution. Finally, $PCF_h$ is derived using Eq. [\ref{eq:main}].

\begin{figure}[H]
    \centering
    \includegraphics[width=0.8\textwidth]{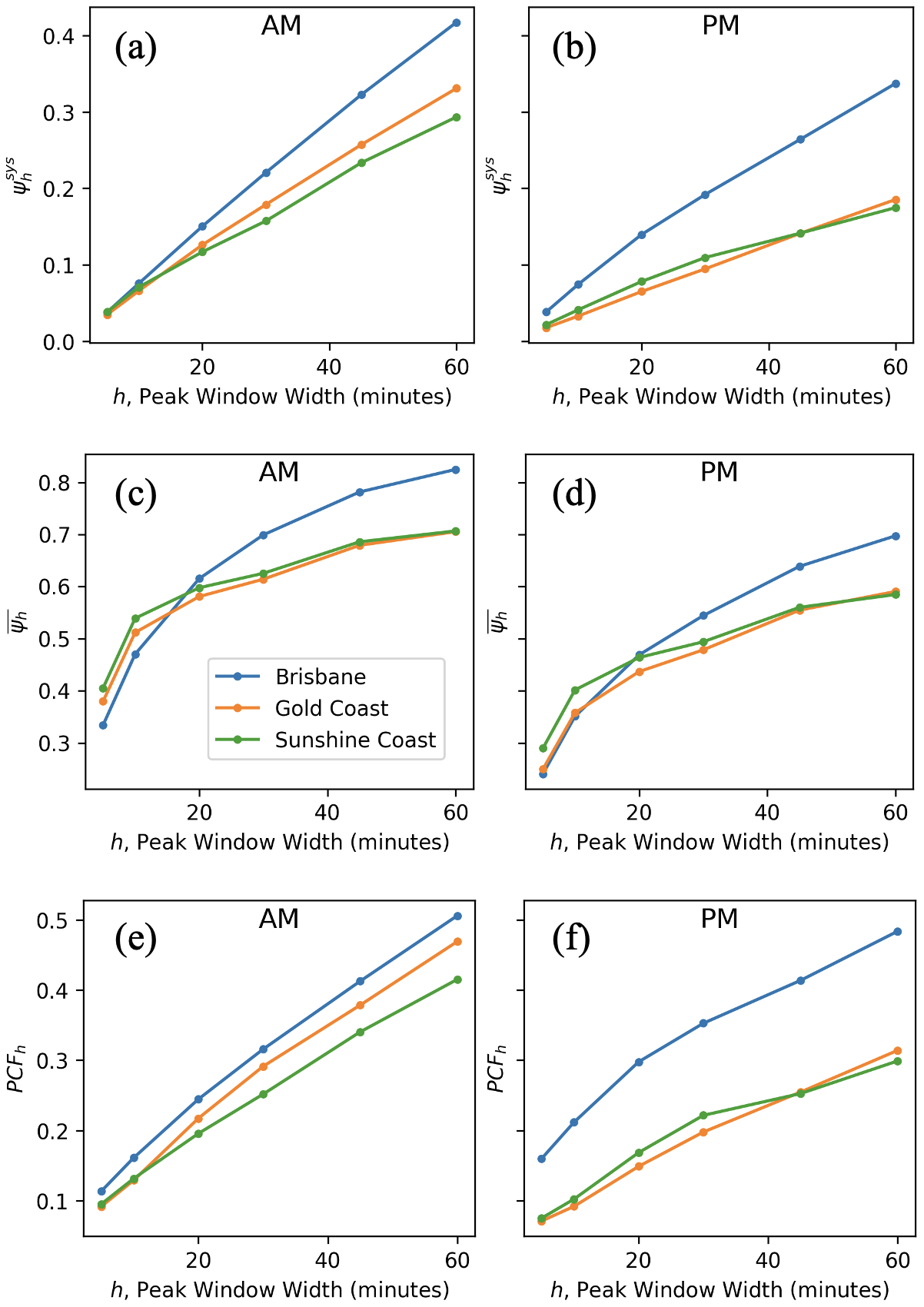}
    \caption{The peakedness characteristics of the departure time distributions for bus commuters from Brisbane, Gold Coast, and Sunshine Coast: [\textbf{a, b}] system-wide departure time peakedness $ \psi_h^{sys} $, [\textbf{c, d}] user-level departure time peakedness $ \overline{\psi_h} $, and [\textbf{e, f}] inter-user peak trip coincidence $PCF_h$, measured at different values of peak window width $h$ for the AM and PM periods. }
    \label{fig:line_plot}
\end{figure}

Figure \ref{fig:line_plot} shows these estimates for different peak window widths, $ h \in \{ 5, 10, 20, 30, 45, 60 \}$. Across the cities, Brisbane exhibits overall higher $ \overline{\psi_h} $ and higher $ PCF_h $, and thus higher $ \psi_h^{sys} $, than Gold Coast and Sunshine Coast in both the AM and PM periods. This is likely due to Brisbane's higher population density and greater urbanisation, which create more structured, work-driven commuting patterns, resulting in more synchronised and concentrated departure times. On the other hand, the more leisure-oriented nature of Gold Coast and Sunshine Coast tends to result in more flexible and dispersed departure patterns. Between the AM and PM periods, peakedness measures are overall higher during the AM period than during the PM period in each city. This can be attributed to the stricter time constraints in the morning, with commuters needing to adhere to fixed start times for work or school, leading to a greater concentration of AM departure time distributions around their peaks. However, Brisbane's $PCF_h$ measures display an interesting variation: $PCF_h$ of PM trips is higher than or similar to that of AM trips, indicating that users' peak departure time windows coincide more strongly in the PM period than in the AM period, while the other two cities show much lower peak coincidence in the PM period. This can be explained by a more uniform end-of-day schedule across work sectors and after-school activities in Brisbane, resulting in greater synchronisation of departure times across individual users.

\subsection{Estimation of Optimal Peak Window}\label{sec:results_optimal_h}

\begin{figure}[htbp]
    \centering
    \begin{subfigure}{0.65\textwidth}
        \includegraphics[width=\textwidth]{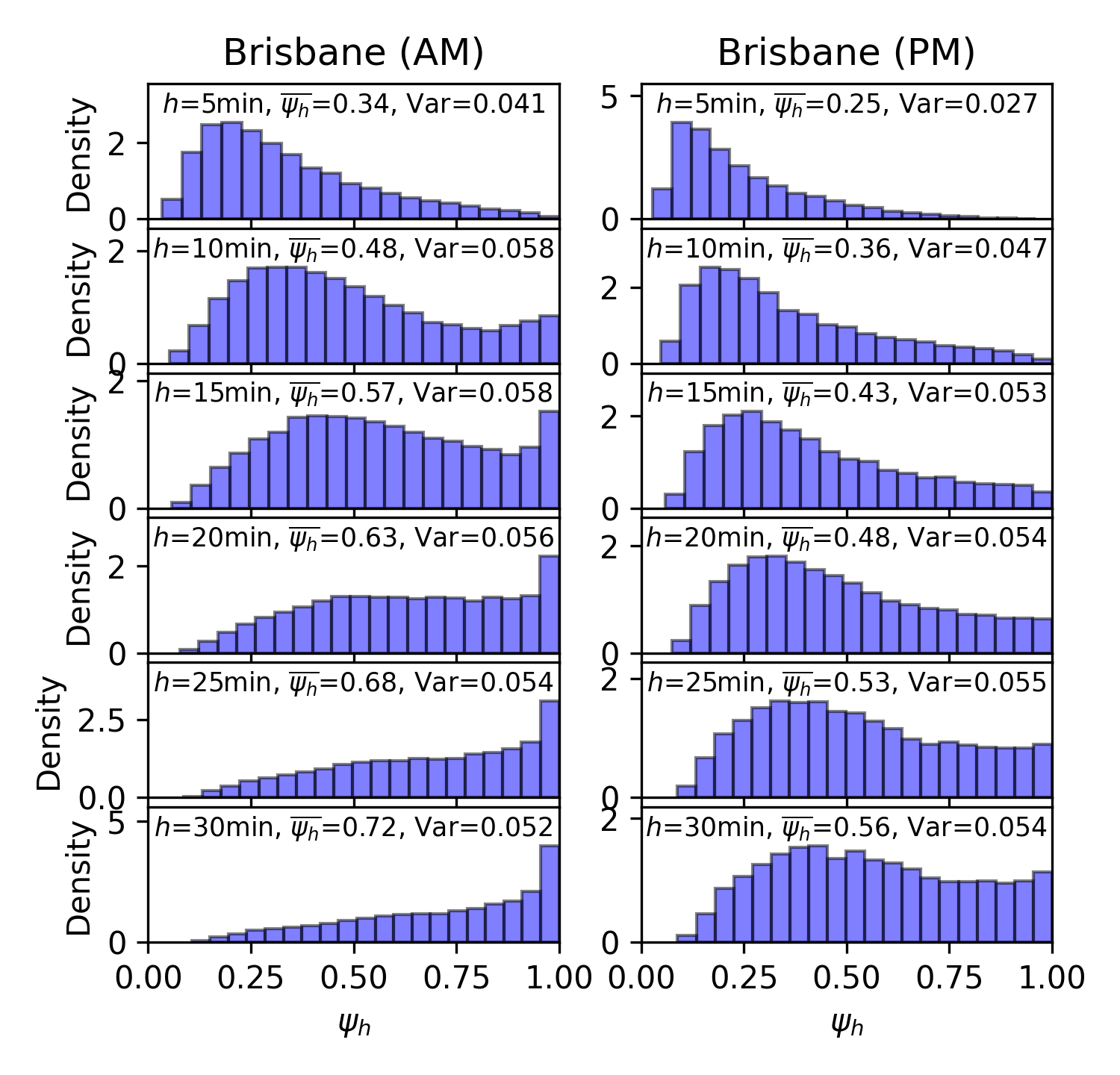}
        \caption{Brisbane}
        \label{fig:hist_bne}
    \end{subfigure}
    
    \begin{subfigure}{0.65\textwidth}
        \includegraphics[width=\textwidth]{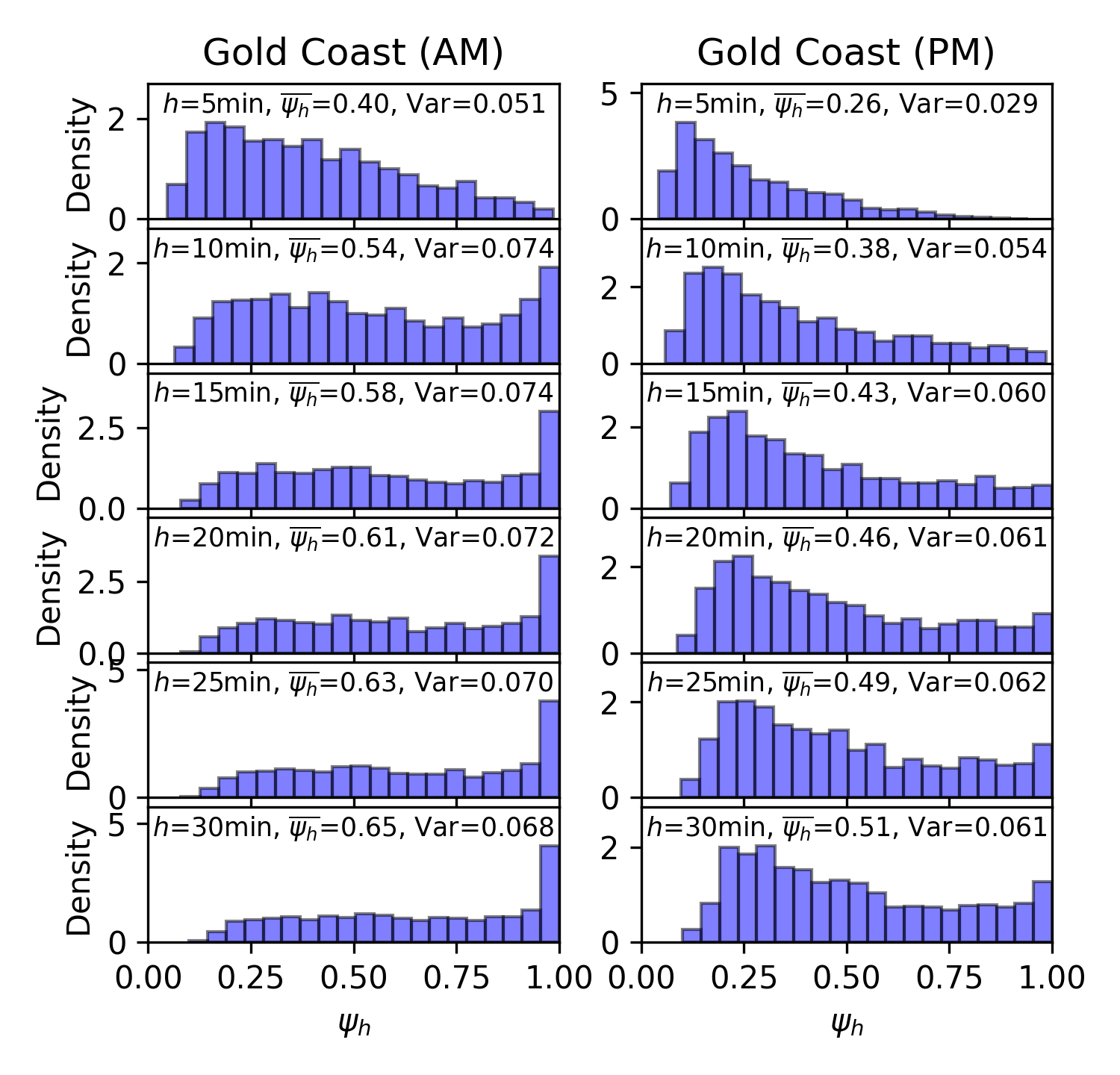}
        \caption{Gold Coast}
        \label{fig:hist_gc}
    \end{subfigure}
    
    \caption{Histograms of estimated $ h $-min peak trip concentration ($ \psi_h $) across users under different $ h $ values ($ h=5, 10, \cdots, 30 $ minutes): the mean ($ \overline{\psi_h} $) and variance (Var) of $\psi_h$ are presented in each histogram.}
    \label{fig:hist_hpp}
\end{figure}

\begin{figure}[h]
    \includegraphics[width=\textwidth]{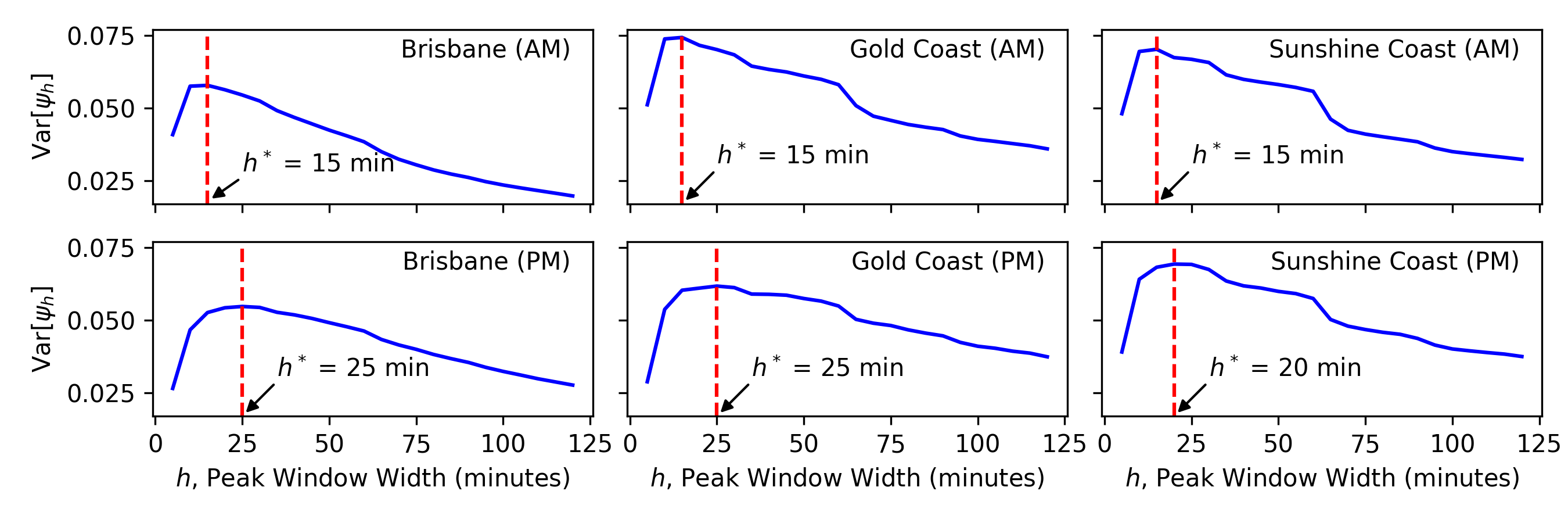}
    \caption{The variance ($\text{Var}[\psi_h]$) of estimated $ h $-min peak concentration ($ \psi_h $) across users under different $ h $ values. The $ h $ value that maximises the variance (i.e., optimal peak window, $ h^* $) is marked by a vertical dashed line.}
    \label{fig:line_opt_h}
\end{figure}

Figure~\ref{fig:hist_hpp} shows the estimation results of $ \psi_h $ for different $h$ values, namely $ h = 5, 10, \cdots, 30 $ minutes, for Brisbane and Gold Coast during two time-of-day periods (AM and PM). For each $h$, the distribution of the measured $ \psi_h $ values across users is shown as a histogram, along with its mean ($ \overline{\psi_h} $) and variance (Var). As expected, the distribution of the measured $ \psi_h $ values shift to the right (i.e., the mean, $ \overline{\psi_h} $, increases) as the window width, $h$, becomes larger. Meanwhile, the diversity or spread of the measured $ \psi_h $ values (i.e., $\text{Var}[\psi_h]$) initially increases and then decreases as $h$ increases. Figure~\ref{fig:line_opt_h} provides a separate plot showing the change in $\text{Var}[\psi_h]$ with respect to $ h $, where the $ h $ value that maximises the variance (optimal peak window, $ h^* $) is marked by a vertical dashed line.

The analysis indicates that the value of $ h^* $ is consistent across the three study networks. Specifically, all three network indicate that $ h^* $ is 15 minutes for the AM period and $ h^* $ is 20-25 minutes for the PM period. The shorter optimal window for the AM period likely reflects more concentrated departure time distributions, as morning schedules are typically more constrained due to fixed start times for work or school. In contrast, departure times may be more spread out due to the varying end times of activities. When many individuals have tightly concentrated departure times (as in the AM period), the variance of $\psi_h$ values across users will peak at a shorter $h$ because smaller windows are sufficient to capture the high concentration of departures for most users. On the other hand, when the population has more dispersed or spread-out departure times (as in the PM period), the variance of $\psi_h$ across the population will reach its peak at a larger $h^*$ value because a larger $h$ window is needed to capture significant portions of the departure events across users. Despite the differences in the optimal window values between the AM and PM periods, the variance around those values is not significantly different. Overall, the analysis for Brisbane, Gold Coast, and Sunshine Coast reveals that the optimal window $ h^* $ is consistently around 20 minutes, regardless of the time-of-day group or region. This suggests that there is a common temporal threshold where departure time peakedness is most distinguishable across the population. Thus, we consider using $h$=20 minutes as the most effective window for measuring trip concentrations ($ \psi_h $), as it reveals the greatest diversity in departure time peakedness.

Table~\ref{tab:user_stat} presents the descriptive statistics of the distribution of estimated $ \psi_h $ values with $h$=20 minutes (i.e., $\psi_{20}$). Across the three cities, on average, approximately 61-63\% of daily departure times occur during each user's most concentrated 20-minute window in the AM period, and 46-49\% in the PM period. Users in the top 5\% for peakedness, represented by the 95th percentile, show that nearly 98-100\% of their departures are concentrated within the peak 20-minute window in the AM and 92-97\% in the PM. In contrast, users in the bottom 5\% for peakedness, as indicated by the 5th percentile, have only 20-23\% of their departure times within the peak 20-minute window for the AM period, and 16-17\% for the PM period.

Understanding the optimal window $ h^* $ and the resulting measurement distribution can serve as a guide for identifying users with high departure time peakedness more accurately. Practically, this offer empirical evidence for transport planners to design targeted policies for those with high trip concentration in the 20-minute window (e.g., those with 20-minute peak trip concentration, $ \psi_{20} $, above the 75th percentile) as these users represent groups with high departure time rigidity and therefore more likely face higher penalties in the event of service disruptions. Additionally, this distribution allows us to determine how peaked a specific user is. For example, a Brisbane traveller with a $ \psi_{20} $ of 0.5 (i.e., 50\% of daily departures occur within the peak 20-minute window) would be considered relatively low in peakedness for the AM period as it is below the median of 0.638 in Table~\ref{tab:user_stat}, but relatively high for the PM period as it is above the median of 0.442.

\begin{table}[t!]
    \centering
    \caption{Descriptive statistics of individual user peak trip concentration ($\psi_h$) across users (with $ h $ = 20 minutes)}

    \begin{tabular}{lrrrrrr}
    \toprule
     & \multicolumn{2}{c}{Brisbane} & \multicolumn{2}{c}{Gold Coast} & \multicolumn{2}{c}{Sunshine Coast}\\
     & AM & PM & AM & PM & AM & PM \\
    \midrule
    Count & 23505 &	23120 &	1834 &	1577 &	795 &	720 \\
    Mean & 0.632 & 0.484 & 0.614 & 0.462 & 0.632 & 0.493 \\
    StdDev & 0.237 & 0.233 & 0.268 & 0.247 & 0.260 & 0.263 \\
    P5 & 0.235 & 0.167 & 0.198 & 0.160 & 0.203 & 0.159 \\
    P25 & 0.443 & 0.294 & 0.377 & 0.255 & 0.410 & 0.264 \\
    Median & 0.638 & 0.442 & 0.604 & 0.403 & 0.645 & 0.440 \\
    P75 & 0.839 & 0.654 & 0.870 & 0.644 & 0.870 & 0.691 \\
    P95 & 0.985 & 0.922 & 1.000 & 0.934 & 1.000 & 0.974 \\
    \bottomrule
    \multicolumn{7}{l}{\small StdDev: standard deviation, Px: x-th percentile}
    \end{tabular}
    
    \label{tab:user_stat}
\end{table}

\subsection{Factors Affecting Departure Time Peakedness}\label{sec:results_factors}
We investigated factors affecting departure time peakedness by grouping the data by specific factors and comparing the three peakedness measures ($ \psi_h^{sys} $, $ \overline{\psi_h} $, and $PCF_h$) across the groups for the three city networks (Brisbane, Gold Coast, and Sunshine Coast) and the two time-of-day periods (AM and PM). A peak window width of $h$=20 minutes is used for the peakedness measurement. 

\begin{figure}[H]
    \centering
    \includegraphics[width=\textwidth]{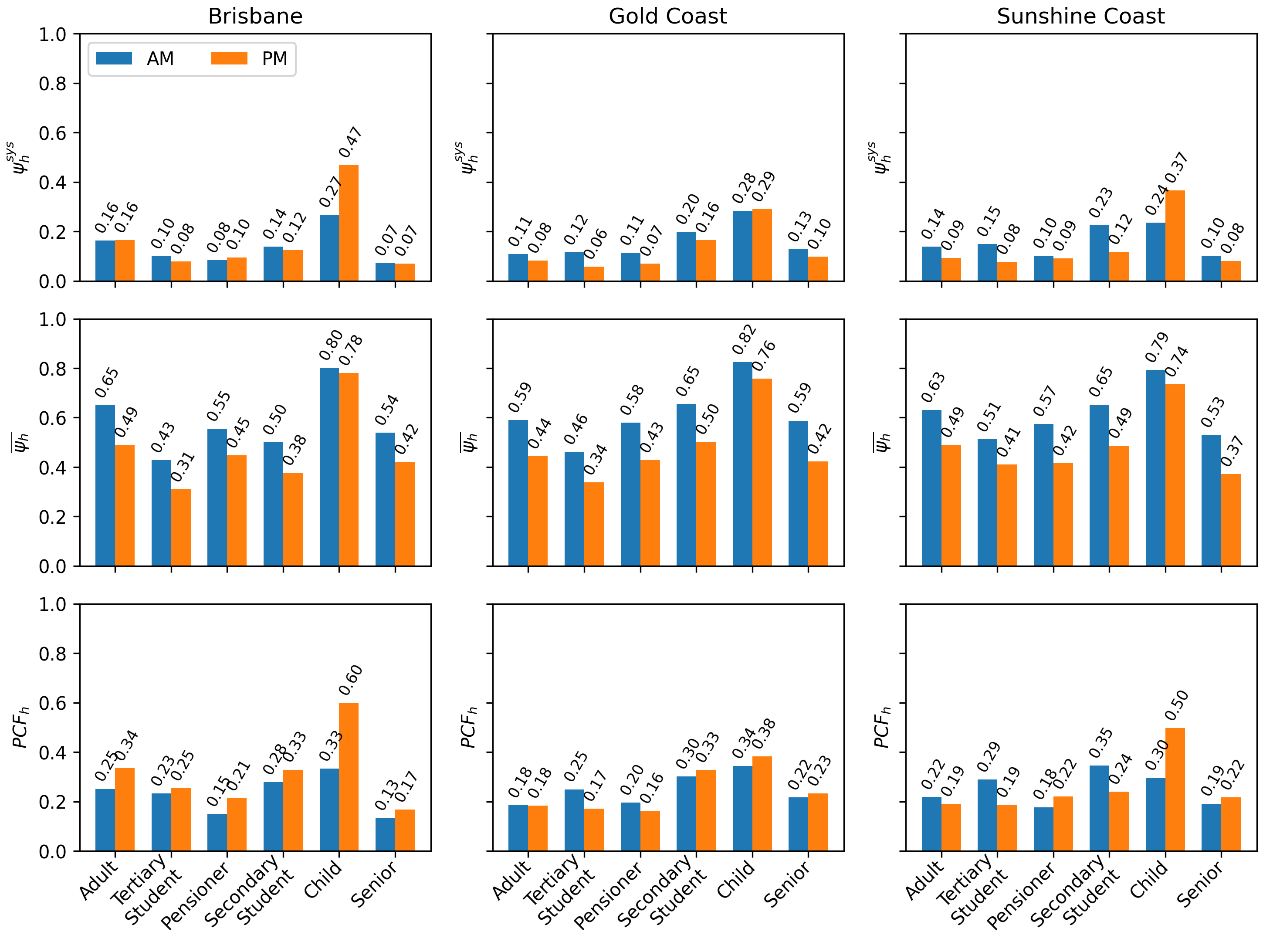}
    \caption{Comparison of peakedness measures ($ h $ = 20 minutes) across different passenger types}
    \label{fig:bar_plot_passenger}
\end{figure}

First, we explored how different passenger types shape departure time peakedness. Figure~\ref{fig:bar_plot_passenger} compares the three peakedness measures across six passenger types: Adult, Tertiary Student, Pensioner, Secondary Student (15+ years), Child (5 to 14 years), and Senior. Notably, the child passenger type exhibits consistently high departure time peakedness across all three measures, particularly in user-level peakedness ($ \overline{\psi_h} $). For example, in Brisbane, the $ \overline{\psi_h} $ of children is nearly 0.8, indicating that most of their day-to-day departures occur within their most frequent 20-minute departure time window. This high concentration is likely due to the fixed start and end times of primary and secondary school, which children in the 5-14 year age group typically attend. Interestingly, the peak window coincidence ($PCF_h$) for the PM period is notably high for children in Brisbane ($PCF_h$=0.6) and Sunshine Coast ($PCF_h$=0.5), even surpassing the AM period. This could be attributed to the more uniform scheduling of after-school activities or programs across various areas in these cities, which may lead to a higher alignment of peak departure times among different users within the peak window. This finding provides some evidence to the role of deeply ingrained travel habits in shaping the travel choice of children \citep{mucelli2016} alongside other factors including service availability, perceived safety and accessibility. 

\begin{figure}[H]
    \centering
    \includegraphics[width=\textwidth]{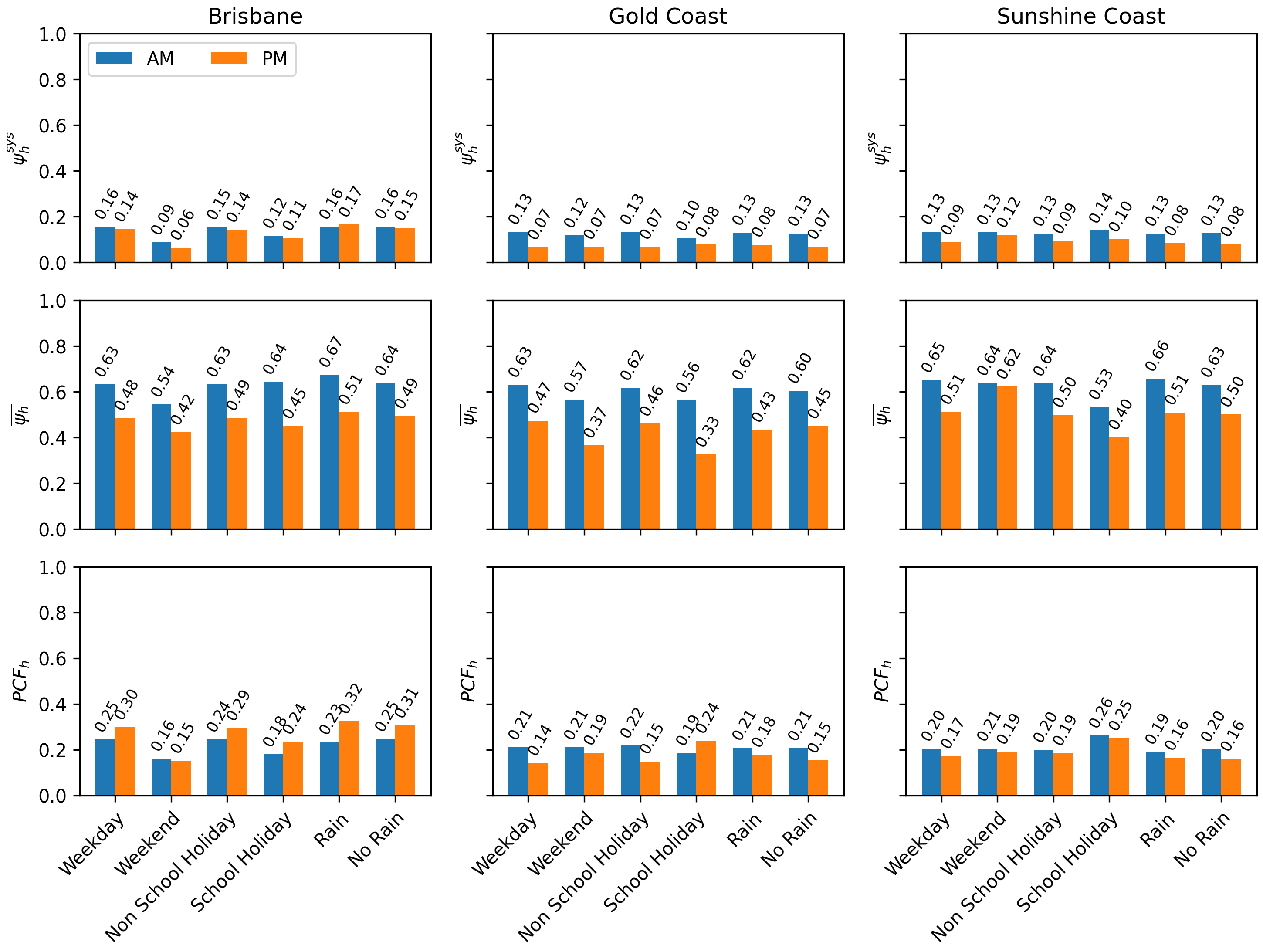}
    \caption{Comparison of peakedness measures ($ h $ = 20 minutes) under different scenarios}
    \label{fig:bar_plot_scenario}
\end{figure}

Next, Figure~\ref{fig:bar_plot_scenario} draws together a suite of situational variables known to impact bus ridership dynamics in the form of six scenarios: Weekday vs Weekend, Non School Holiday vs. School Holiday, and Rain vs. No Rain. We find that the departure time peakedness characteristics are relatively stable across these scenarios, with some subtle shifts. For example, in Brisbane and Gold Coast, weekend departure times are slightly less peaked than weekdays, likely due to more flexible weekend schedules with diverse non-work activities, leading to more dispersed travel times. In contrast, weekdays are dominated by structured work and school schedules, resulting in higher peakedness. Sunshine Coast does not show such distinction, possibly due to its tourism-driven, leisure-oriented travel patterns and flexible local employment, resulting in consistent travel behaviour across both weekdays and weekends.

Overall, the analysis of both passenger type comparisons and scenario factors reveals that peakedness measures vary more significantly across different passenger types, while remaining relatively stable across situational variables. This suggests that departure time peakedness is more deeply tied to inherent, passenger-specific characteristics rather than external situational factors. It supports the idea that individuals’ habitual travel behaviours play a dominant role in shaping their departure time choices, with these ingrained patterns showing less sensitivity to external conditions like weather, holidays, or day type.

\subsection{Temporal Stability and Variability in Departure Time Peakedness: A Trend Analysis}\label{sec:results_trend_analysis}
Next we investigate long-term temporal dynamics of departure time peakedness to understand whether system- or user-level peakedness characteristics change over time. To explore this, we measured both $\psi_h^{sys}$ and $\psi_h$, with $ h $ = 20 minutes, using data from a three-month moving window spanning over one year from July 2015 to June 2016. This results in a time-series with 10 data points, corresponding to the periods July-September 2015, August-October 2015, and so on, up to April-June 2016. These data points allow us to analyse trends or changes in the chosen peakedness measure over time.

For the \textit{user-level} analysis, a time-series of $\psi_h$ was constructed for each user-OD pair. To ensure statistical reliability, we required a minimum of 12 observations (equivalent to at least one trip per week) in each three-month window for valid $\psi_h$ estimates. Users meeting this criterion, i.e., those making trips consistently throughout the entire year, are referred to as `\textit{long-term} regular users'. These users form a subset of the `regular users' identified for this study, as shown in Table \ref{table:datasets}. Table \ref{table:trend} shows the number of long-term regular users for each city, along with the original count of regular users and the percentage of long-term regular users. In Brisbane, long-term regular users account for approximately 40\% of the original regular user base, while in Gold Coast and Sunshine Coast, they represent about 20\%. For the \textit{system-level} analysis, a time-series of $\psi_h^{sys}$ was constructed for each city by combining departure time distributions of all associated long-term regular users. We estimated $\psi_h^{sys}$ for each three-month moving window, conducting separate analyses for the AM and PM periods.

\begin{table}
	\centering
	\caption{Counts of user-OD pairs for long-term regular users and original regular user base}
	\label{table:trend} 
	\vspace{-3mm}
	\includegraphics[width=0.9\textwidth]{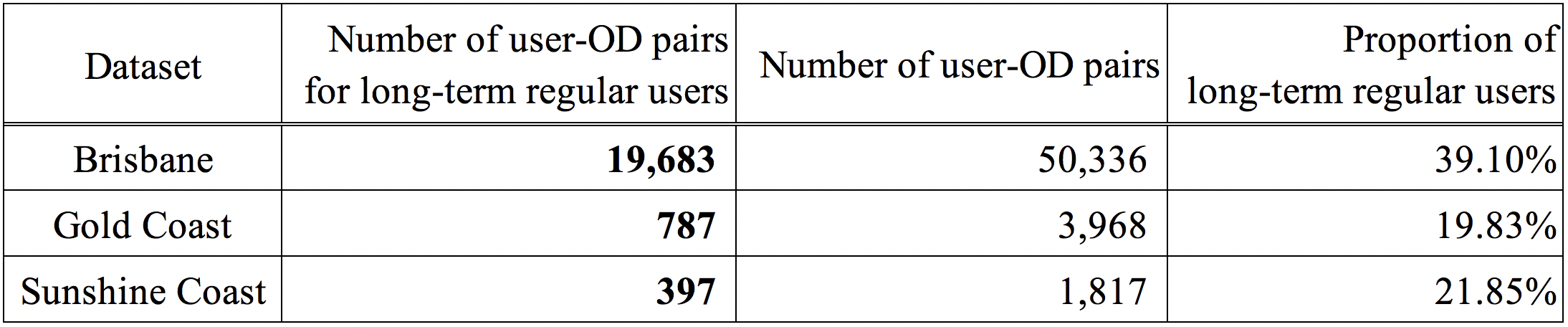}\\
\end{table}

To evaluate the presence of trends, we applied the Mann-Kendall (MK) Trend Test \citep{mann1945nonparametric, kendall1948rank}, which assesses whether there is a statistically significant monotonic increase or decrease in the peakedness measures over time. As our time series data are relatively short, we used the original non-parametric MK test for analysing consistently increasing or decreasing trends (monotonic trends), which does not consider serial correlation or seasonality effects \cite{hussain2019pyMannKendall}. The MK test determines whether to reject the null hypothesis of no monotonic trend at a given significance level $\alpha$, which is set to 0.05 in this study.

\begin{figure} [h]
	\centering
	\includegraphics[width=\textwidth]{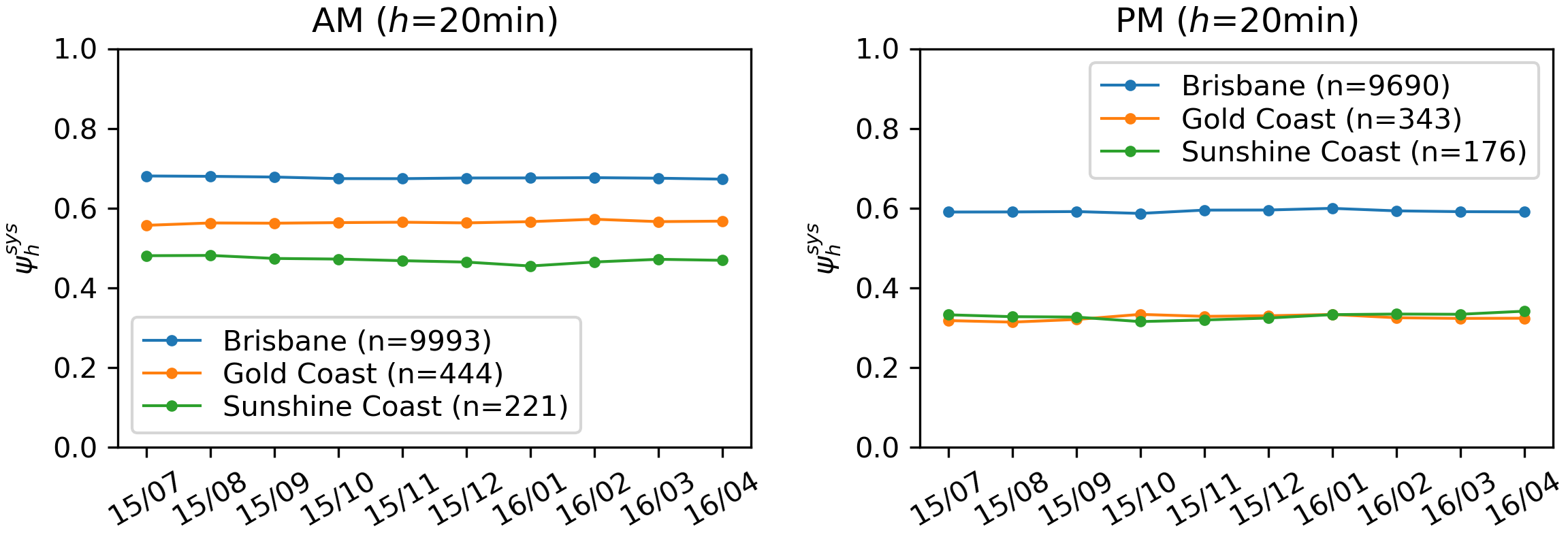}\\
	\caption{Estimates of system-level departure time peakedness ($\psi_h^{sys}$) over time: X-axis shows the first month of each 3-month estimation window (YY/MM)}
	\label{fig:line_trend_sp}
\end{figure}

For the system-level peakedness, the MK test results show that there is no monotonic trend in the series, which is true for all three cities for both AM and PM. Figure \ref{fig:line_trend_sp} shows the time-series of $\psi_h^{sys}$ used in the MK test, where the x-axis labels  the first month of each three-month moving window in the format `YY/MM' (Year/Month). The number of long-term regular users involved in constructing each system-level time-series ($n$) is shown in the parenthesis next to each city name within the legend. In all six cases, the value of $\psi_h^{sys}$ remains quite stable over the one-year period, indicating the absence of a significant trend in the system-level departure time peakedness. 

\begin{figure} [h]
	\centering
	\includegraphics[width=\textwidth]{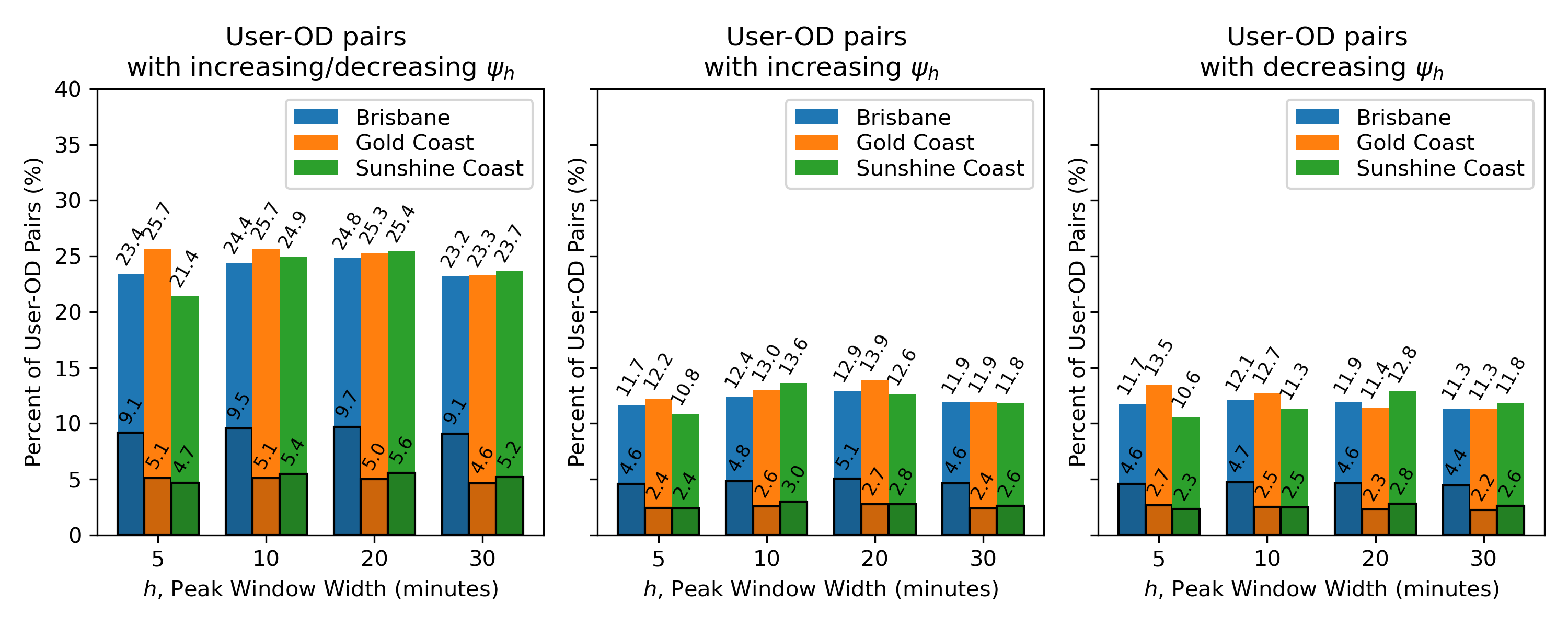}\\
	\caption{User-OD pairs with increasing or decreasing $\psi_h$, identified for three cities and four peak windows: values are shown as a percentage of \textit{long-term regular users} (indicated by taller bars in the background) and as a percentage of the original \textit{regular user base} (indicated by shorter bars in the foreground)}
	\label{fig:bar_trend_percent}
\end{figure}

In contrast, the MK test results reveal interesting trends in the user-level peakedness time-series data. Approximately 25\% of the long-term regular users exhibit statistically significant trends in their $\psi_h$ time-series. Of these users, half show increasing trends and the other half show decreasing trends. This pattern is consistent across all three cities, regardless of the selected peak window width ($ h $). The detailed results are displayed in Figure \ref{fig:bar_trend_percent}, which shows the proportion of user-OD pairs with decreasing or increasing $\psi_h$, both as a percentage of \textit{long-term regular users} (indicated by taller bars in the background) and as a percentage of the original \textit{regular user base} (indicated by shorter bars in the foreground). The results are obtained for the three cities using four different peak window widths ($ h $ = 5, 10, 20, and 30 minutes). The first sub-figure shows the number of user-OD pairs exhibiting any monotonic trends (either decreasing or increasing $\psi_h$), while the second and third sub-figures show the numbers corresponding to increasing and decreasing trends, respectively.

\begin{figure} [h]
	\centering
	\includegraphics[width=\textwidth]{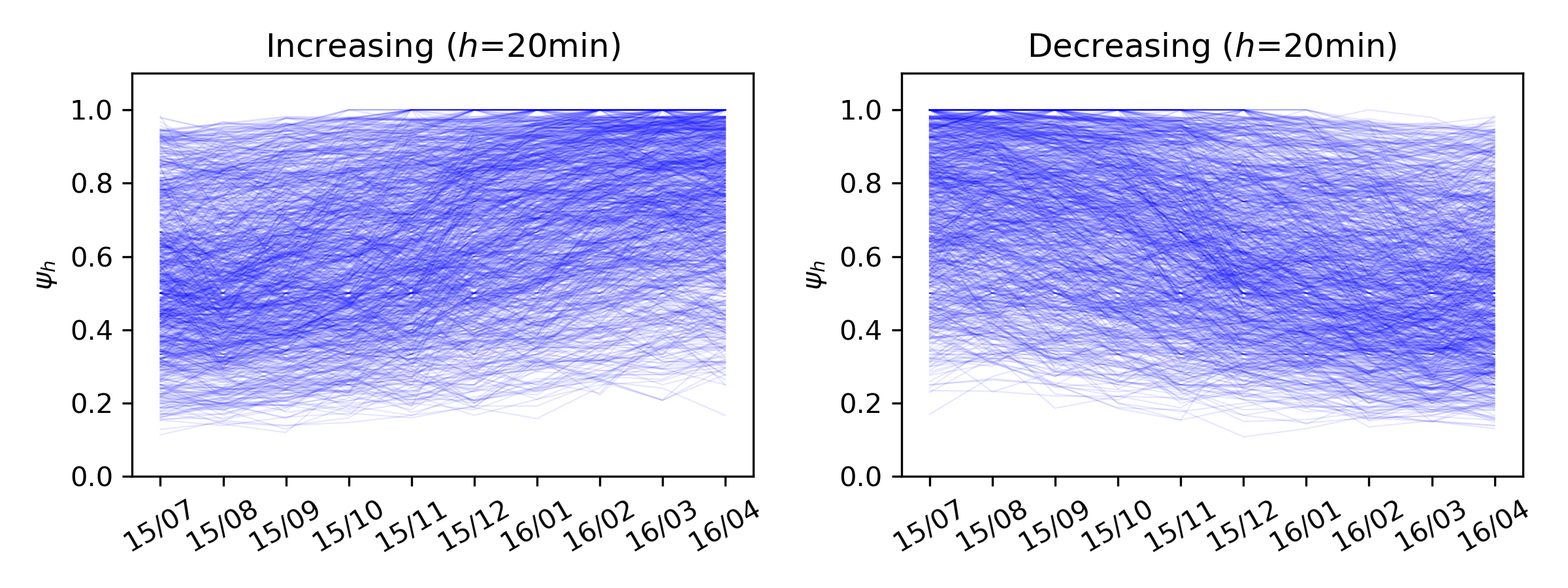}\\
	\caption{Time-series of $\psi_h$ for the long-term regular users with monotonic trends (Brisbane)}
	\label{fig:line_trend_user}
\end{figure}

Figure \ref{fig:line_trend_user} shows the time-series of $\psi_h$ for the long-term regular users with monotonic trends. The users with increasing trends and decreasing trends are first separated and 1,000 cases are randomly sampled from each group and displayed on the left and right figures, respectively. The $\psi_h$ measures are from the Brisbane data set, based on $ h $ = 20 minutes. A visual examination of the two plots reveals some interesting patterns. For example, for increasing trends a step change during the period 15/11 to 15/12 that temporally aligns to a comparable step change for users with a decreasing trend. This finding suggests that shifts in service availability and scheduling may have each acted to create this outcome. 

\begin{figure} [h]
	\centering
	\includegraphics[width=\textwidth]{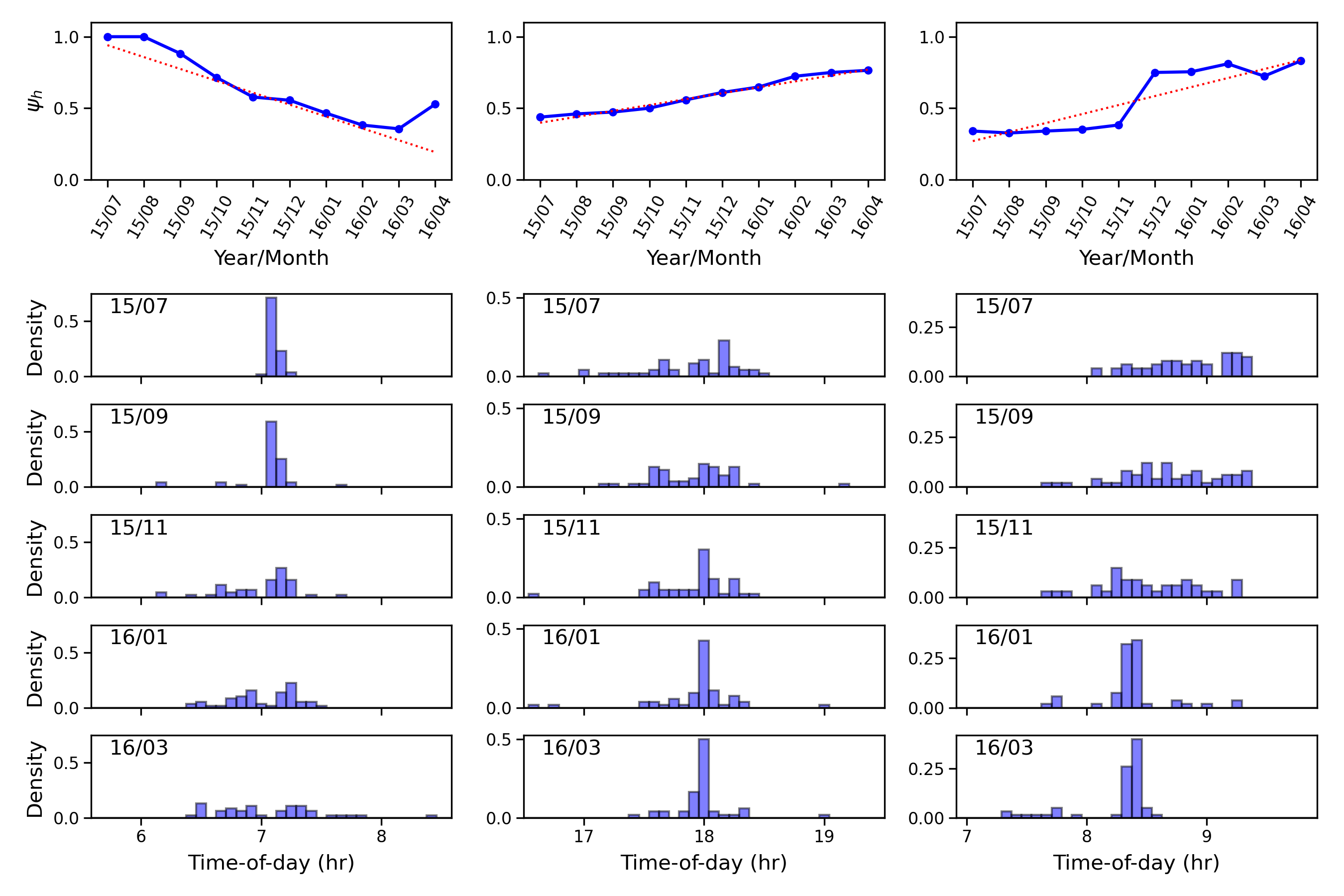}\\
	\caption{Long-term changes in departure time peakedness: examples of three selected users}
	\label{fig:user_hpp_trend_example}
\end{figure}

Figure \ref{fig:user_hpp_trend_example} provides examples of three selected users, whose departure time peakedness change over time. Each column is associated with one user, where the top plot shows the time-series of $\psi_h$ over one year and the histograms below show the underlying three-month departure time distribution that resulted in each $\psi_h$ value for five selected observation points (every two months along the year). As depicted by the first case, a user's departure time distribution can be highly peaked at the beginning and then gradually dispersed over time, producing the pattern of decreasing $\psi_h$. The opposite pattern is also observed, where the departure time distribution is initially dispersed but become concentrated around a single peak, leading to increasing $\psi_h$ as in the second case. Some users show more sudden changes, rather gradual increase or decrease in peakedness, as depicted by the third case. In this case, the value of $\psi_h$ remains at 0.3-0.4 throughout the first five months and then jumps to 0.75-0.8 by exhibiting a clear shift in the peakedness of the departure time distribution. This finding goes some way to reveal how a series of factors (that are beyond the scope of the current paper and indeed the go card dataset) seemingly extraneous to the individual traveller such as service availability (both increases and decreases) has created the conditions under which shifts in departure time peakedness have taken place. A follow up study that was designed to follow a set of long-term users and delineate the extraneous forces shaping shifts in departure time peakedness would be extremely valuable.  

To sum, at the system level, our analysis results show no significant monotonic trends in any city (Brisbane, Gold Coast, or Sunshine Coast), indicating that system-level departure time peakedness remains stable over time. This suggests that the collective travel behaviour of long-term regular users does not exhibit major fluctuations in departure time patterns throughout the year. However, at the user level, the analysis reveals notable dynamics. Approximately 25\% of long-term regular users (those making trips consistently throughout the year) exhibit significant trends in their $\psi_h$ time-series. Interestingly, these trends are balanced: about half of the users show increasing peakedness (with more concentrated and regular departure times over time), while the other half display decreasing peakedness (with more dispersed and flexible travel patterns over time). Some users exhibit gradual changes, while others experience sudden shifts. Overall, this trend analysis highlights the stability of system-level peakedness, while revealing the evolving nature of individual-level departure time behaviour. It also suggests that habitual travel patterns, though resistant to short-term situational factors like weather or holidays (as shown in Section \ref{sec:results_factors}), may shift over the long term due to larger-scale factors such as seasonality, service changes, or shifts in user characteristics.

\subsection{Applications of Peakedness Measures}\label{sec:results_impact}

One of the most critical applications of our peakedness measurement framework is in \textit{disruption management}. Commuters with higher levels of departure time peakedness (represented as higher user-level $ \psi_h $ values) would tend to have higher rigidity and lower flexibility in departure time choice and, thus, have higher rescheduling penalties and lower willingness to shift departure time. As such, service disruptions that require departure time changes can cause higher impact on users with stronger departure time peakedness than on those with lower peakedness levels. Traditional transit planning and operations often rely on system-level departure time distributions to identify peak hours and assess demand patterns. However, these aggregated patterns do not necessarily reflect individual user behaviours, particularly their flexibility or rigidity in departure time choices.

Our method reveals that user-level peakedness and system-level peakedness do not always exhibit a direct correlation. In some cases, areas with users whose departure time distributions have a high degree of peakedness may still result in a combined departure time distribution at the system level that shows low peakedness. This discrepancy arises because the system-level distribution peakedness ($ \psi_h^{sys} $) is influenced by both the peakedness of individual departure time distributions ($ \overline{\psi_h} $) and the alignment of their peak time windows ($ PCF_h $). In situations where the peak windows of individuals are widely dispersed (resulting in very low $ PCF_h $), the system's peakedness can remain low, even when user-level peakedness is high (as illustrated in Figure \ref{fig:mixture_dist}(c)-(d). Assessing the disruption impacts based only on aggregated system-level departure distributions would not accurately capture the rescheduling penalties faced by individual users. In order to provide more user-centric transit services and improve user experience in managing disruptions, it is important to assess what a given disruption means to each user by considering individual users' departure time peakedness characteristics.

\begin{figure}[htbp]
    \centering
    \includegraphics[width=0.9\textwidth]{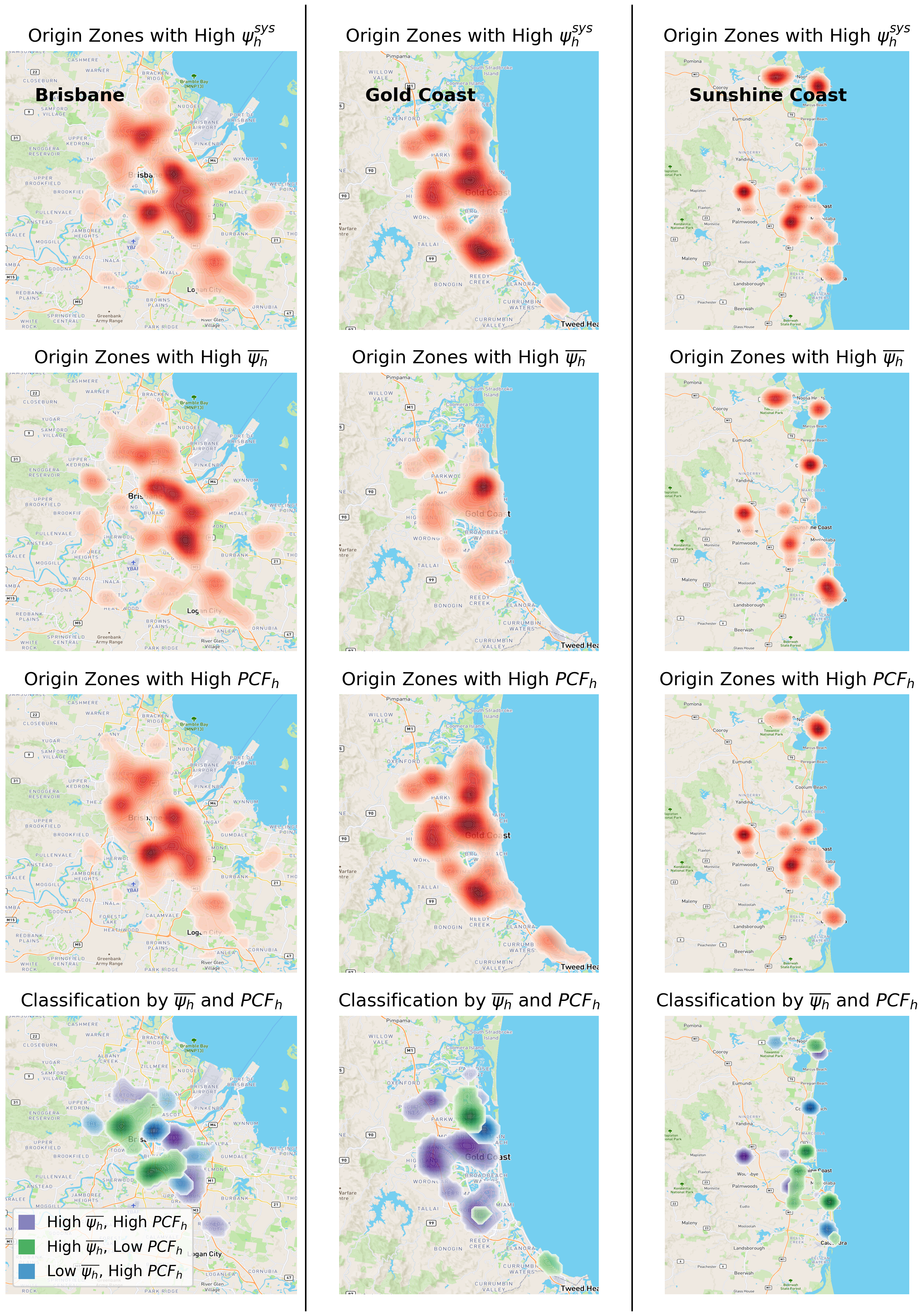}
    \caption{The spatial distribution of origin zones with high departure-time peakedness in terms of three peakedness measures, namely $ \psi_h^{sys} $, $ \overline{\psi_h} $, and $ PCF_h $ (AM period).}
    \label{fig:zone_contour_city}
\end{figure}

To demonstrate how the proposed peakedness measurement framework can assist public transit operators in understanding the potential impact of disruptions on transit users, we estimate the three peakedness measures for each origin zone, defined as the 1-km radius circle centred around each bus stop. We identify origin zones with high peakedness in terms of each measure type, namely $ \psi_h^{sys} $, $ \overline{\psi_h} $, or $ PCF_h $, within each of the three cities (Brisbane, Gold Coast, and Sunshine Coast) during the AM period. For a given measure type, we obtain the values of that peakedness measure for all origin zones within each city and determine the median value. Origin zones with peakedness measures greater than this median are considered to have high peakedness for that measure type. Figure \ref{fig:zone_contour_city} shows the spatial distribution of the origin zones with high peakedness in terms of $ \psi_h^{sys} $, $ \overline{\psi_h} $, and $ PCF_h $ in the first, second, and third rows, respectively. Notably, there are considerable discrepancies between origin zones with high system-wide departure time peakedness ($ \psi_h^{sys} $) and origin zones with high user-level departure time peakedness ($ \overline{\psi_h} $). This observation empirically validates our conjecture that system-wide demand patterns do not consistently reflect their individual users' departure time choice patterns. We argue that it is more informative to examine the two user-level components ($ \overline{\psi_h} $ and $ PCF_h $) separately, rather than relying on the aggregated measure ($ \psi_h^{sys} $), as demonstrated in the fourth row of Figure \ref{fig:zone_contour_city}. For instance, transit operators can prioritise origin zones with both high $ \overline{\psi_h} $ and high $ PCF_h $. Users in these zones exhibit higher departure time peakedness, leading to increased rescheduling penalties during service disruptions, and strong peak window coincidence, meaning more users will be affected if a service disruption occurs during the system-wide peak period. By prioritising these zones, transit operators can proactively address the challenges posed by disorders, minimising the overall impact on users and the whole system. In addition, transit operators may develop tailored strategies for different scenarios. For instance, in origin zones with high $ \overline{\psi_h} $ and low $ PCF_h $, users may exhibit high departure time peakedness individually, but their peak travel times are not as tightly clustered. Prioritising these zones might be beneficial in specific situations, such as disruptions that affect a broader time window. On the other hand, origin zones with low $ \overline{\psi_h} $ and high $ PCF_h $ could be prioritised when the goal is to minimise the impact on a large number of users within a specific time frame, even if their departure time distributions are more dispersed. Ultimately, the choice of prioritisation strategy among these scenarios depends on the nature of the service disruption and the specific performance goals that transit operators aim to achieve.

We have also conducted an analysis investigating destination zones to provide a more comprehensive understanding. This additional analysis, presented in \ref{si:dest_zones}, examines arrival time peakedness and reveals similar discrepancies between system-wide and user-level peakedness measures as observed in the origin-based analysis. By considering both origin and destination zones, transit operators can identify areas with high peakedness in both departure and arrival patterns, allowing them to develop more tailored disruption management strategies that account for rigid travel behaviours at both ends of the journey.

\section{Conclusion}
Our individual mobility habits are critical to our transport system implicating congestion, environmental pollution, and the travel experience. There is currently no empirical measure that can capture individual mobility habits at scale. The current study implements the first such measure capturing the habitual aspects of day-to-day departure time distribution. Measuring the time distribution of public transit riders' tendency towards choosing the same departure time repeatedly is important in its capacity to provide the empirical measure of travel habits of individual users across a metropolis. Our measure of departure time peakedness comprises a set of three related measures that collectively capture a user's tendency towards repeatedly choosing the same departure time for a given origin and destination within the bus network. High departure time peakedness indicates users `sticking' to a preferred departure time, with low peakedness capturing users with less preference towards a particular time of departure. 

Our findings demonstrate that departure time peakedness is for the main part driven by intrinsic passengers characteristics (captured in the current study by passenger type: adults, students, children, seniors and pensioners), rather than external variables such as weather and holiday periods. Furthermore, while individual passengers show significant variation in their departure time peakedness over extended periods of time (offering some evidence to suggest that personal travel habits may transform gradually), at the overall system-level, peakedness exhibits remarkably long-term stability. Another key insight from our analysis is that system-level departure time peakedness does not always reflect individual user-level patterns. Traditional transit planning and operations often rely on system-level departure time distributions to identify peak hours and assess demand patterns. However, this information alone does not offer insights into individual travel behaviours, such as users' responses towards changes or disruptions in public transport services. Distinguishing user-level peakedness from system-level peakedness is crucial for assessing disruption impacts and developing targeted mitigation strategies.
For public transit operators, considering individual users' departure time peakedness is important in understanding the relative impact of a service changes/disruptions. Given that users associated with high departure time peakedness are typically associated with higher rescheduling penalties and lower willingness to shift departure time, service changes/disruptions creating departure time changes will likely to impact users with stronger departure time peakedness than on those with lower peakedness levels. 

From a managerial perspective, the choice of the peak window width $h$ (in the current study that was optimally defined at 20 minutes) has important implications. For example, a smaller $h$ (e.g., 5 minutes) allows for the identification of extremely rigid travellers. Since the proportion of trips occurring within such a narrow window is generally small, users with high $h$-min peak concentrations within this small window indicate a very rigid and consistent departure time preference. Conversely, a larger $h$ (e.g., 30 minutes) allows us to identify particularly flexible travellers. Since trip proportions within a larger window are generally higher, users with low $h$-min peak concentration in this context can be considered as having particularly more dispersed departure time patterns, potentially less vulnerable to service disruptions. For strategic planning, a smaller $h$ might help focus on the most intense peaks, while a larger $h$ would have value in informing boarder capacity planning decisions. In demand management, exploring changes in peak trip concentration before and after a major intervention---such as the introduction of flexible working arrangements by a major employer, the commencement of congestion charging or the introduction of highly subsided fares (as seen in SEQ)---can reveal the extent to which policies resulted in increasing or redistributing demand or simply shifting the peak window. From an operational perspective, the choice of the peak window width could be aligned to operational capabilities, such that for urban transit, the 20-minute window (that we employed in this study) might be most appropriate, whereas for other part of the system (such as the airport transit services) or for weekend operations a larger window width may work better.

For metropolitan councils and policy makers, understanding how broader changes in land use, transport policy and pandemics act to re-shape users' departure time peakedness are important in helping to inform future initiatives. This might include the likely implication on users following the removal, modification or introduction of new routes and services. Particular stops and routes associated with higher levels of peakedness are likely to induce more disruption if shifts to service scheduling and routes were imposed compared to stops and routes with users characterised by lower peakedness levels.

Considering whether high or low departure time peakedness is a positive or negative characteristic is dependent on both the situation and who we are concerned about. For individual users, higher levels of peakedness may be an indicator of limited choice (in terms of fulfilling a given daily mobility goal, such as transiting from home to work) thus a potentially negative characteristic. For transit operators, metropolitan councils and policy makers, higher levels of peakedness (associated with a given stop or route) are generally a good feature in terms of setting and deploying the requisite resources (such as service frequency and capacity) to reliably meet demand. In planning for future introduction of new routes and services, targeting these where the user base is highly peaky might `unstick' the user base through offering more `choice' would likely result in positive outcomes. Conversely, avoiding the removal/modification of routes and services whose user base are highly peaky would be important to minimise disruptions to their user base.

Our framework can be extended in various ways to further explore habitual travel behaviour and its implications. First, while we focus on smart card data for bus passengers, the peakedness measurement can be applied to any departure or arrival time data, such as Global Positioning System (GPS) records, to analyse broader traveller populations, including private cars and freight. This could reveal different departure or arrival time peakedness patterns across modes with varying levels of flexibility and constraints. Second, integrating richer activity, demographic, and socio-economic data---such as trip purpose, income, occupation, or household structure---could provide deeper insights into how trip context and personal circumstances influence habitual travel behaviours. Finally, studying how departure time peakedness changes in response to major disruptions (e.g., pandemics, extreme weather, or infrastructure changes) could enhance our understanding of travel behaviour resilience and adaptation. A critical implication is that enabling such investigations requires longitudinal day-to-day observations of repeated trip choices over an extended period to capture proper choice distributions. Single-day travel data, like traditional household travel surveys, are insufficient for in-depth analysis of habitual behaviours. Expanding access to longitudinal mobility data through smart card systems, GPS, Internet-of-Things (IoT) devices, and other emerging technologies is essential for advancing research in this area.

The capacity to measure individual travel habits at scale is an essential prerequisite to understanding the causes of congestion, pollution and the factors that shape the travel experience. In an era defined by climate change and the shift toward sustainable mobility, such tools have never been more urgent. It is only with such empirical tools that smarter public transit systems can be designed in a manner that best meet the evolving needs of citizen's everyday mobility requirements.

\bibliographystyle{elsarticle-harv} 
\bibliography{main}






\clearpage
\appendix

\section{Limitations of Existing Measures} \label{si:existing_measures}
While several existing measures can characterise the concentration of a distribution, they may not fully capture the specific aspect of `peakedness' we are interested in here. Common measures such as the \textbf{\textit{standard deviation}} or \textbf{\textit{variance}} describe the overall spread of a distribution, but they fail to focus on how observations are clustered around a central peak. These measures of dispersion capture the extent to which the entire distribution is stretched or squeezed, including the influence of extreme values in the tails. However, the peakedness we aim to measure emphasises the configuration of data near the peak, rather than the overall spread. This is important because users with high peakedness—those whose departure times are consistently clustered around a preferred time—may still exhibit occasional deviations. Traditional dispersion measures, like standard deviation, are highly sensitive to these outliers and may misrepresent a user's typical departure behaviour, making them unsuitable for capturing the regularity we seek to quantify.

\textbf{\textit{Kurtosis}} is another common measure that describes the shape of a distribution, particularly its `tailedness' and the sharpness of its peak. While it provides some insight into whether a distribution has a sharp or flat peak, the interpretation of kurtosis in terms of peakedness is incorrect because it does not directly measure the shape of the peak (`peakedness')---the regularity or consistency of repeated observations around the peak)---but primarily describes the shape of the tails (`tailedness') \citep{westfall2014kurtosis}. Moreover, kurtosis is sensitive to outliers and extreme values in the tails of the distribution, which can inflate or distort the measure even if the central concentration of data remains strong. 

Another potential measure, the \textbf{\textit{entropy}} of the distribution, assesses the unpredictability or uniformity of choices, yet it lacks a direct focus on the presence and intensity of a peak. While entropy can give insight into the overall variability of departure time choices, it does not specifically focus on how consistently observations cluster around a peak. Entropy treats all variations equally, meaning it does not differentiate between a user whose departure times are highly concentrated around a preferred time and one whose choices are spread across multiple periods, so long as both exhibit the same level of unpredictability. Moreover, entropy is a global measure, considering the entire range of the distribution, and does not prioritise local structures such as a central peak. 

In conclusion, existing measures such as standard deviation, kurtosis, and entropy fail to adequately isolate the regularity or concentration of repeated observations around a specific departure time. Our proposed `peakedness' measure is designed to address this gap, focusing on how sharply and consistently a user's departure time choices align with a preferred peak. Unlike traditional metrics, our measure is robust and not sensitive to outliers, as it quantifies the proportion of observations concentrated around the peak. Regardless of how far an outlier may deviate from the peak, the peakedness remains unaffected as long as the central concentration of departure times remains consistent. This makes our measure more reliable for capturing the true regularity in users' departure time choices, providing a clearer and more focused understanding of their day-to-day behavioural patterns.

\section{Birnbaum's Concept of Relative Peakedness}\label{si:relative_peakedness}
In his 1948 paper `On Random Variables with Comparable Peakedness' \cite{birnbaum1948random}, Birnbaum introduced the concept of `relative peakedness' to correctly measure what is intuitively considered as the peakedness of a distribution and evaluate the concentration of probability distributions around their peaks. The key idea behind relative peakedness is to quantify how sharply or strongly a distribution is concentrated around its mode or peak 'relative to' other distributions. That is, it assesses how much one distribution is more `peaked' compared to another reference distribution, rather than evaluating the peakedness of a distribution in isolation. Birnbaum's definition of relative peakedness is presented below: 
\begin{definition}[Relative Peakedness \cite{birnbaum1948random}]
	\label{def:birnbaum}
	Let $ Y $ and $ Z $ be real random variables and $ Y_1 $ and $ Z_1 $ real constants. We shall say that $ Y $ is more peaked about $ Y_1 $ than $ Z $ is about $ Z_1 $ if the inequality
	\begin{equation} \label{eq:rel_peakedness}
	P(|Y-Y_1| \geq T) \leq P(|Z-Z_1| \geq T)
	\end{equation}
	is true for all $ T\geq 0 $.
\end{definition}

Birnbaum’s definition of relative peakedness provides a way to compare the concentration of two random variables, $ Y $ and $ Z $, around their respective reference points $ Y_1 $ and $ Z_1 $. To make this comparison clearer, Eq.(\ref{eq:rel_peakedness}) can be rewritten as $ P(Y_1-T < Y < Y_1+T) \geq P(Z_1-T < Z < Z_1+T) $, meaning that if the distribution of $ Y $  is more peaked than that of $ Z $, the probability of $ Y $ being within a $ \pm T $ window of its reference point $ Y_1 $ is greater than or equal to the probability of $ Z $ being within the same distance $ T $ from $ Z_1 $.  In other words, if $ Y $ is more peaked around $ Y_1 $ than $ Z $ is around $ Z_1 $, then $ Y $ is more tightly concentrated around $ Y_1 $, exhibiting less dispersion around $ Y_1 $ compared to $ Z $ around $ Z_1 $, for any given distance $ T $ ($ T\geq 0 $).

A limitation of Birnbaum’s relative peakedness is that it only establishes a comparative condition between two distributions and does not provide a direct measure to quantify the peakedness of an individual distribution. To address this limitation, we extend Birnbaum's concept to develop a quantitative measure that characterises the peakedness of a single probability distribution. Inspired by the idea of assessing the peakedness of a distribution based on `the concentration of data points within a specified range', we measure the degree to which data points are concentrated within the most peaked range of an individual distribution, focusing on `the highest concentration of data points', rather than relative concentration of data points as in Birnbaum's approach. By applying this concept to travellers' departure time distributions, we propose the \textit{$ h $-minute Peak Trip Concentration} (PTC) metric, $\psi_h$ in Eq. [\ref{eq:psi_h}], which measures how much departure times are concentrated within the most peaked $h$-minute window of the day. This provides a direct measure to quantify the peakedness of an individual departure time distribution.

\section{Conceptual Similarities with Peak Hour Factor}\label{si:peak_hour_factor}
There is  evidence supporting the idea that measuring the concentration of events within the most intense range provides a meaningful way to quantify the peakedness of a distribution. One practical example of this in traffic engineering is the Peak Hour Factor (PHF), which compares the total traffic volume during the peak hour with the traffic volume during the busiest 15-minutes of the peak hour, as follows:

\begin{equation} \label{eq:peak_hour_factor}
PHF = \frac{\text{Total Hourly Traffic Volume}}{\text{Peak 15-minute Traffic Volume} \times 4}
\end{equation}

A lower PHF signifies more pronounced peaks, indicating that traffic is heavily concentrated into the peak 15-minute interval (i.e., sharp demand spikes), while a higher PHF reflects a more even distribution of traffic over the entire peak hour. PHF is widely applied in highway capacity analysis, roadway design, and traffic signal optimisation to better accommodate fluctuations in traffic demand. At its core, PHF captures the proportion of traffic volume concentrated within the most intense 15-minute period of the peak hour. Similarly, the proposed $ h $-minute peak trip concentration, $ \psi_h $, measures the proportion of trips that fall within the most peaked $h$-minute window of the day. While $ \psi_h $ and PHF are defined differently and apply to different domains, they share conceptual similarities in how they quantify peakedness by identifying the interval with the highest concentration of events.

\section{Further Analysis of Optimal Peak Window ($ h^* $)} \label{si:further_analysis_for_h}

\begin{figure}[htbp]
    \centering
    \begin{subfigure}{0.7\textwidth}
    \includegraphics[width=\textwidth]{figures/optimal_h_illust_a.png}
    \caption{}
    \label{fig:optimal_h_illust_a}
    \end{subfigure}\\
    
    \begin{subfigure}{0.7\textwidth}
    \includegraphics[width=\textwidth]{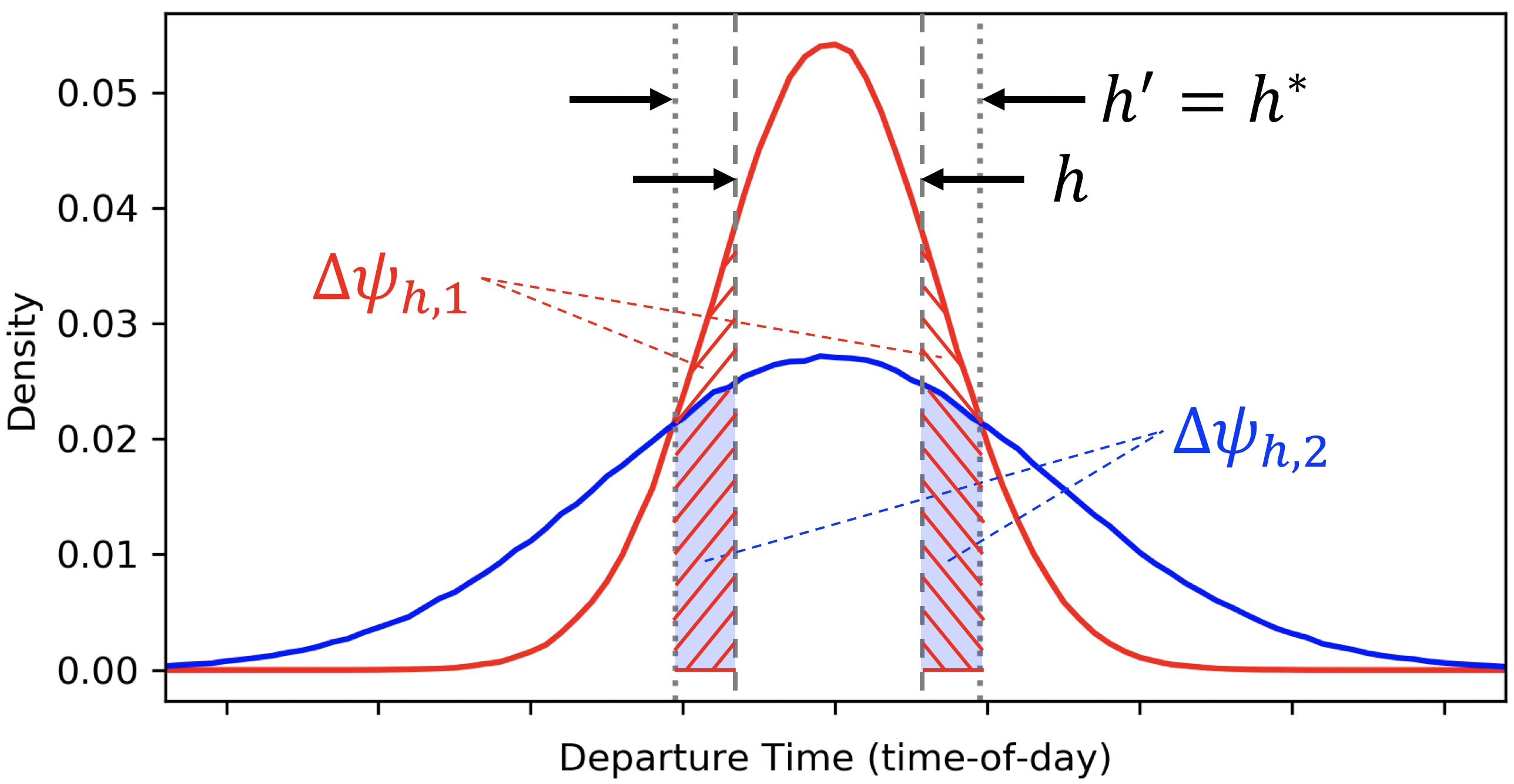}
    \caption{}
    \label{fig:optimal_h_illust_b}
    \end{subfigure} \\
    
    \begin{subfigure}{0.7\textwidth}
    \includegraphics[width=\textwidth]{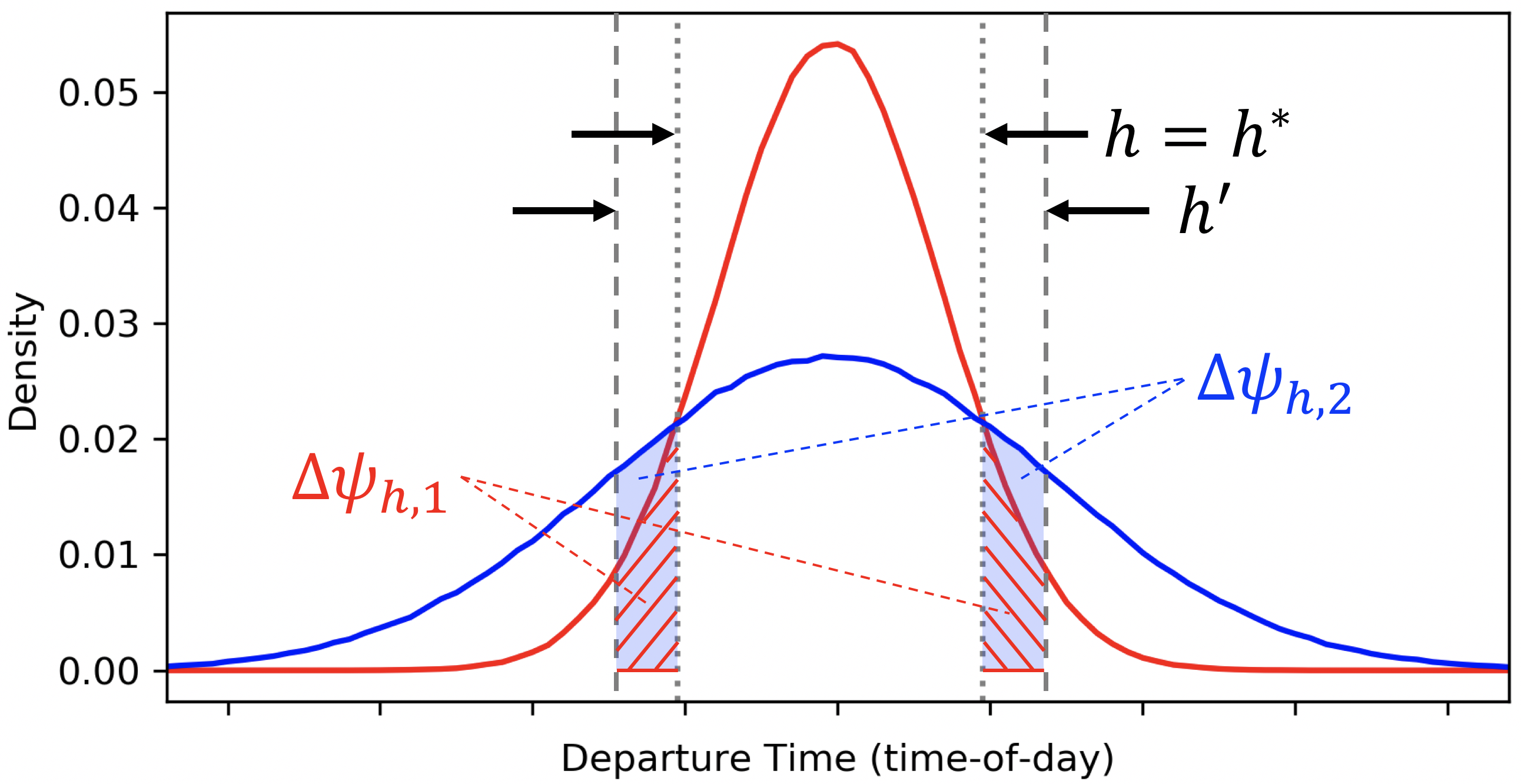}
    \caption{}
    \label{fig:optimal_h_illust_c}
    \end{subfigure}
\caption{Illustration of optimal peak window: the interpretation of variance-maximising $h^*$ in a system with two departure time peakedness levels}
\label{fig:optimal_h_illust} 
\end{figure}

To get some intuition about what the variance-maximising $ h^* $ in Eq.(\ref{eq:optimal_h}) represents, we derive $ h^* $ in a simplified system consisting of two types of users---user group 1 with high peakedness and user group 2 with low peakedness. Figure \ref{fig:optimal_h_illust_a} illustrates the probability density curves for these two user groups, where $ f_1 $ and $ f_2 $ represent the departure time distribution of high peakedness users and low peakedness users, respectively. The $ h $-min peak concentration under $ f_1 $ and $ f_2 $, denoted by $\psi_{h, 1}$ and $\psi_{h, 2}$, are also shown as the shaded areas under the curves. In this setting, we have found that $ h^* $ corresponds to the peak window at which the two density curves intersect as indicated by the two vertical dotted lines in Figure \ref{fig:optimal_h_illust_a}. 

To prove this, we show that the variance of $\psi_h$ increases, i.e., $\text{Var}[\psi_h] < \text{Var}[\psi_{h^{\prime}}]$ for $h < h^{\prime}$, when $h$ and $h^{\prime}$ are less than or equal to $ h^* $, where we show $\text{Var}[\psi_h] < \text{Var}[\psi_{h^{\prime}}]$ occurs only under the condition $ \psi_{h^{\prime}, 1} - \psi_{h, 1} > \psi_{h^{\prime}, 2} - \psi_{h, 2} = \Delta \psi_{h, 1} > \Delta \psi_{h, 2} $ that happens when the $f_1$ curve (red) is above the $f_2$ curve (blue) as illustrated in Figure \ref{fig:optimal_h_illust_b} ($h, h^{\prime} \leq h^* $). Similarly, we show that the variance decreases, i.e., $\text{Var}[\psi_h] \geq \text{Var}[\psi_{h^{\prime}}]$ for $h < h^{\prime}$, when $h$ and $h^{\prime}$ are greater than or equal to $ h^* $, where $\text{Var}[\psi_h] \geq \text{Var}[\psi_{h^{\prime}}]$ occurs only under the condition $ \psi_{h^{\prime}, 1} - \psi_{h, 1} \leq \psi_{h^{\prime}, 2} - \psi_{h, 2} = \Delta \psi_{h, 1} \leq \Delta \psi_{h, 2} $ that happens when the $f_1$ curve (red) is below the $f_2$ curve (blue) as illustrated in Figure \ref{fig:optimal_h_illust_c} ($h, h^{\prime} \geq h^* $). With these, we can see that the variance of $\psi_h$ is maximised at $ h = h^* $, where $h^*$ is the boundary between the two conditions, $\Delta \psi_{h, 1} > \Delta \psi_{h, 2}$ and $\Delta \psi_{h, 1} \leq \Delta \psi_{h, 2}$. This property is formally described by Proposition \ref{prop:optimal_h} below, followed by the detailed proof.

\begin{proposition}
	\label{prop:optimal_h}
	Let  $ h $ and $ h^{\prime} $ be two peak windows such that $ h^{\prime} > h $ and let $ \Delta \psi_{h} $ be the increment in $ h $-min peak concentration by expanding the peak window from $ h $ to $ h^{\prime} $, i.e., $ \Delta \psi_{h} = \psi_{h^{\prime}} - \psi_{h} >0 $. If a system consists of two types of users with unimodal departure time distributions, where user group 1's departure time distribution is more peaked than that of user group 2 such that $ \psi_{h, 1} \geq \psi_{h, 2} $ for all $ h \geq 0 $, then $ h^* $ that maximises the variance of $ \psi_{h} $ across users in the system represents the boundary in $ h $ that satisfies:
	\begin{subequations}
		\begin{align}
		& \Delta \psi_{h, 1} > \Delta \psi_{h, 2}  \quad & \text{for} \quad h <  h^{\prime} \leq h^* \\
		& \Delta \psi_{h, 1} \leq \Delta \psi_{h, 2}  \quad & \text{for} \quad h^*  \leq h < h^{\prime}
		\end{align}
	\end{subequations}
	where $ \Delta \psi_{h, i} = \psi_{h^{\prime}, i} - \psi_{h, i} $ for $ i\in \{1,2\} $.
\end{proposition}

\begin{proof}
	Let $ f_1 $ and $ f_2 $ be the probability densities for departure time distribution of user group 1 (high peakedness) and user group 2 (low peakedness) , respectively, and let $ w_1 $ and $ w_2 $ be the proportions of user group 1 and user group 2 in the system, respectively (i.e., $ w_1 + w_2 = 1 $). Using the following relationship between the variance of $ \psi_{h} $ under $ h $ and $ h^{\prime} $:
	\begin{equation} \label{eq:var_hpp}
	\begin{split}
	\text{Var}[\psi_{h^{\prime}}] &= \text{Var}[\psi_{h} + \Delta \psi_{h}] \\
	&= \text{Var}[\psi_{h}] + \text{Var}[\Delta \psi_{h}] + 2 \, \text{Cov}[\psi_{h}, \Delta \psi_{h}]
	\end{split}
	\end{equation}
	we obtain the expression for $ \text{Var}[\psi_{h^{\prime}}] - \text{Var}[\psi_{h}] $ as follows:
	\begin{equation} \label{eq:var_diff}
	\begin{split}
	\text{Var}[\psi_{h^{\prime}}] - \text{Var}[\psi_{h}] &= \text{Var}[\Delta \psi_{h}] + 2 \, \text{Cov}[\psi_{h}, \Delta \psi_{h}] \\
	&= \sum_{i \in \{1,2 \}} w_i (\Delta \psi_{h,i} - \overline{\Delta \psi_h})^2 + 2 \sum_{i \in \{1,2\}} w_i (\psi_{h, i}-\overline{\psi_h})(\Delta \psi_{h, i} - \overline{\Delta \psi_h}) \\
	&= \left\{ w_1 (\Delta \psi_{h,1} - \overline{\Delta \psi_h})^2 + w_2 (\Delta \psi_{h,2} - \overline{\Delta \psi_h})^2 \right\}  \\
	&\quad + 2 \left\{ w_1 (\psi_{h,1}-\overline{\psi_h})(\Delta \psi_{h, 1} - \overline{\Delta \psi_h}) + w_2 (\psi_{h, 2}-\overline{\psi_h})(\Delta \psi_{h, 2} - \overline{\Delta \psi_h}) \right\}
	\end{split}
	\end{equation}
	where $ \overline{\psi_h} = w_1  \psi_{h,1} + w_2  \psi_{h,2} $ and $ \overline{\Delta \psi_h} = w_1 \Delta \psi_{h,1} + w_2 \Delta \psi_{h,2} $. Let $ \alpha $ denote the difference between the $ h $-min peak concentration increments of user group 1 and user group 2, i.e., $ \alpha = \Delta \psi_{h, 2} - \Delta \psi_{h, 1}  $, where $ \alpha $ may be positive, negative, or zero. Then, we can express $ \overline{\Delta \psi_h} $ as $ \overline{\Delta \psi_h} = w_1 \Delta \psi_{h,1} + w_2 (\Delta \psi_{h,1} + \alpha ) = \Delta \psi_{h,1} + w_2 \alpha $ and obtain the following simplified expressions:
	\begin{equation} \label{eq:delta_psi}
	\begin{split}
	\Delta \psi_{h,1} - \overline{\Delta \psi_h} &= \Delta \psi_{h,1} - (\Delta \psi_{h,1} + w_2 \alpha) =  -w_2 \alpha\\
	\Delta \psi_{h,2} - \overline{\Delta \psi_h} &= \Delta \psi_{h,1} + \alpha - ( \Delta \psi_{h,1} + w_2 \alpha) = (1-w_2) \alpha = w_1 \alpha \\
	\end{split}
	\end{equation}
	Substituting Eq.\ref{eq:delta_psi} to Eq.\ref{eq:var_diff} gives:
	\begin{equation} \label{eq:var_diff_2}
	\begin{split}
	\text{Var}[\psi_{h^{\prime}}] - \text{Var}[\psi_{h}] &= w_1 \, w_2^2 \, \alpha^2 + w_2 \, w_1^2 \, \alpha^2 \\
	&\quad - 2 \, w_1 \, w_2 \, \alpha \, (\psi_{h, 1}-\overline{\psi_h}) + 2 \, w_1 \, w_2 \, \alpha \, (\psi_{h, 2}-\overline{\psi_h}) \\
	&= w_1 \, w_2 \, \alpha^2 \, (w_1 + w_2) - 2 \, w_1 \, w_2 \, \alpha \, (\psi_{h, 1} - \psi_{h, 2}) \\
	&= w_1 \, w_2 \, \alpha \, \{ \alpha - 2 \, (\psi_{h, 1} - \psi_{h, 2}) \} 
	\end{split}
	\end{equation}
	From Eq.\ref{eq:var_diff_2}, we can derive the following conditional statements:
	\begin{subequations}
		\begin{align}
		& \text{If} \quad \alpha < 0 \quad \text{or} \quad \alpha > 2 \, (\psi_{h, 1} - \psi_{h, 2}),  \quad & \text{then} \quad \text{Var}[\psi_{h^{\prime}}] > \text{Var}[\psi_{h}]  \label{eq:var_greater} \\
		& \text{If}  \quad 0 \leq \alpha \leq 2 \, (\psi_{h, 1} - \psi_{h, 2}),  \quad & \text{then} \quad \text{Var}[\psi_{h^{\prime}}] \leq \text{Var}[\psi_{h}] \label{eq:var_less}
		\end{align}
	\end{subequations}
	which is independent of the user group proportions, $ w_1 $ and $ w_2 $. It is noted that the condition  $ \alpha > 2 \, (\psi_{h, 1} - \psi_{h, 2}) $ is not possible because, if we assume $ \alpha = \Delta \psi_{h, 2} - \Delta \psi_{h, 1} > 2 \, (\psi_{h, 1} - \psi_{h, 2}) $ is true, then 
	\begin{subequations}
		\begin{align}
		\Delta \psi_{h, 2} - \Delta \psi_{h, 1} &> 2 \, (\psi_{h, 1} - \psi_{h, 2}) \\ 
		(\Delta \psi_{h, 2} - \Delta \psi_{h, 1} ) - (\psi_{h, 1} - \psi_{h, 2}) &> \psi_{h, 1} - \psi_{h, 2} \\
		(\Delta \psi_{h, 2} +  \psi_{h, 2}) - (\Delta \psi_{h, 1} + \psi_{h, 1}) &> \psi_{h, 1} - \psi_{h, 2} \\
		\psi_{h^{\prime}, 2} - \psi_{h^{\prime}, 1} &> \psi_{h, 1} - \psi_{h, 2} > 0 \\
		\psi_{h^{\prime}, 2} - \psi_{h^{\prime}, 1} &> 0
		\end{align}
	\end{subequations}
	results in a contradiction since we know that $ \psi_{h^{\prime}, 1} \geq \psi_{h^{\prime}, 2} $ as $ f_1 $ is more peaked than $ f_2 $. Therefore, Eq.\ref{eq:var_greater} and Eq.\ref{eq:var_less} can be simplified to:
	\begin{subequations}
		\begin{align}
		& \text{If} \quad \Delta \psi_{h, 1} > \Delta \psi_{h, 2},  \quad & \text{then} \quad \text{Var}[\psi_{h^{\prime}}] > \text{Var}[\psi_{h}]  \label{eq:var_greater_2} \\
		& \text{If}  \quad \Delta \psi_{h, 1} \leq \Delta \psi_{h, 2},  \quad & \text{then} \quad \text{Var}[\psi_{h^{\prime}}] \leq \text{Var}[\psi_{h}] \label{eq:var_less_2}
		\end{align}
	\end{subequations}
	At $ h=0 $, we have $ \Delta \psi_{h, 1} > \Delta \psi_{h, 2} $. Then, the condition changes from Eq.\ref{eq:var_greater_2} to Eq.\ref{eq:var_less_2} as $ h $ increases. As such, $ h $ at the boundary of this condition change maximises $ \text{Var}[\psi_{h}] $ and, thus, the variance-maximising optimal $ h^* $ satisfies the following conditions:
	\begin{subequations}
		\begin{align*}
		& \Delta \psi_{h, 1} > \Delta \psi_{h, 2}  \quad & \text{for} \quad h <  h^{\prime} \leq h^* \\
		& \Delta \psi_{h, 1} \leq \Delta \psi_{h, 2}  \quad & \text{for} \quad h^*  \leq h < h^{\prime}
		\end{align*}
	\end{subequations}
	
	It is noted that this Proposition is independent of the proportions of user groups---it only depends on the shape of $ f_1 $ and $ f_2 $---, meaning that $ h^* $ remains the same regardless of the group proportions as long as the two distributions, $ f_1 $ and $ f_2 $, are unchanged. For instance, $ h^* $ would be the same when the percentages of user group 1 and user group 2 are (10\%, 90\%), (50\%, 50\%), or (80\%, 20\%). This Proposition also holds when $ f_1 $ or $ f_2 $ is an asymmetric distribution.
	
\end{proof}
Although Proposition \ref{prop:optimal_h} describes a simplified system with only two types of departure time distributions, it gives a useful intuition that the variance-maximising optimal $ h^* $ captures the best peak window that naturally separate different peakedness levels as much as possible, which offers a good middle ground for $h$ that is not too small and not large given the set of departure time distributions in the dataset.

\section{Departure Time Peakedness vs. Arrival Time Peakedness} \label{si:dep_vs_arr}
\begin{figure} [htbp]
	\centering
	\includegraphics[width=.7\textwidth]{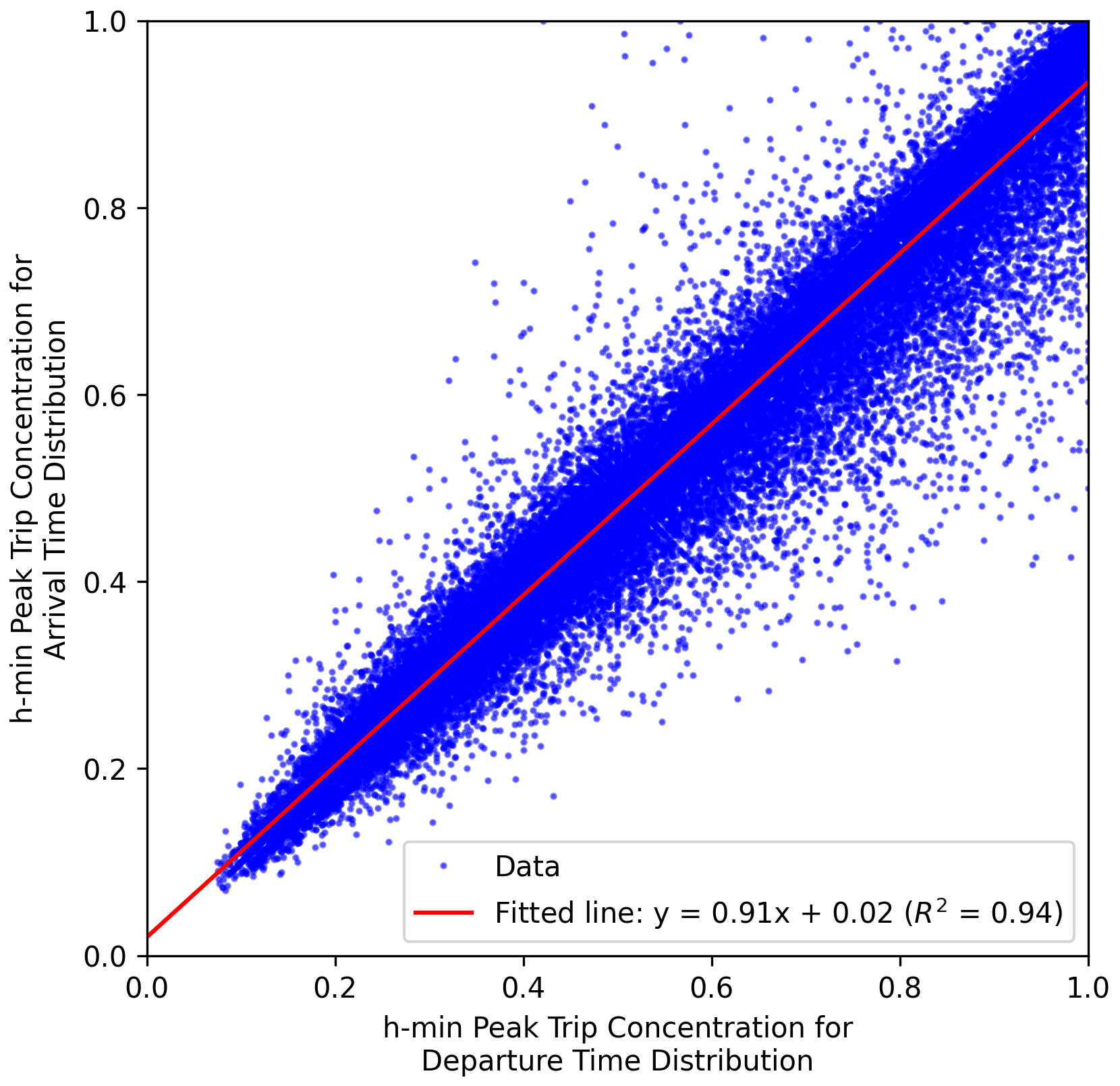}\\
	\caption{A scatter plot of $ h $-minute Peak Trip Concentration, $ \psi_h $, for departure time vs. arrival time distributions for individual users in Brisbane ($h$=20 minutes). The strong correlation between departure time peakedness and arrival time peakedness is demonstrated by the nearly 45-degree linear fit line (red).}
	\label{fig:scatter_dep_arr}
\end{figure}

An analysis was conducted to compare how the peakedness measure of a user may differ between departure times at the origin and arrival times at the destination. For each individual user in the Brisbane dataset, the $ h $-minute Peak Trip Concentration, $ \psi_h $, was estimated based on both the departure time distribution and the arrival time distribution for a given origin-destination (OD) trip using $h$=20 minutes. Figure \ref{fig:scatter_dep_arr} shows a scatter plot where each blue dot represents a pair of $ \psi_h $ values---one from the departure time distribution and the other from the arrival time distribution for the same user. The results indicate a strong positive correlation between departure time peakedness and arrival time peakedness, as evidenced by the nearly 45-degree fitted linear regression line (shown in red): $y=0.91+0.02$ ($R^2=0.94$). This finding suggests that, despite additional factors affecting arrival times such as congestion, travel time variations, and route changes, the degree of peakedness in arrival times remains largely consistent with that of departure times. In other words, users who exhibit high departure time peakedness also tend to have have high arrival time peakedness. This supports the validity and robustness of using departure time peakedness as a meaningful indicator of individual travel routines to study habitual behaviours, without requiring to explicitly factor in arrival distributions.

\clearpage
\section{Destination Zones with High Arrival Time Peakedness} \label{si:dest_zones}

\begin{figure}[htbp]
    \centering
    \includegraphics[width=0.6\textwidth]{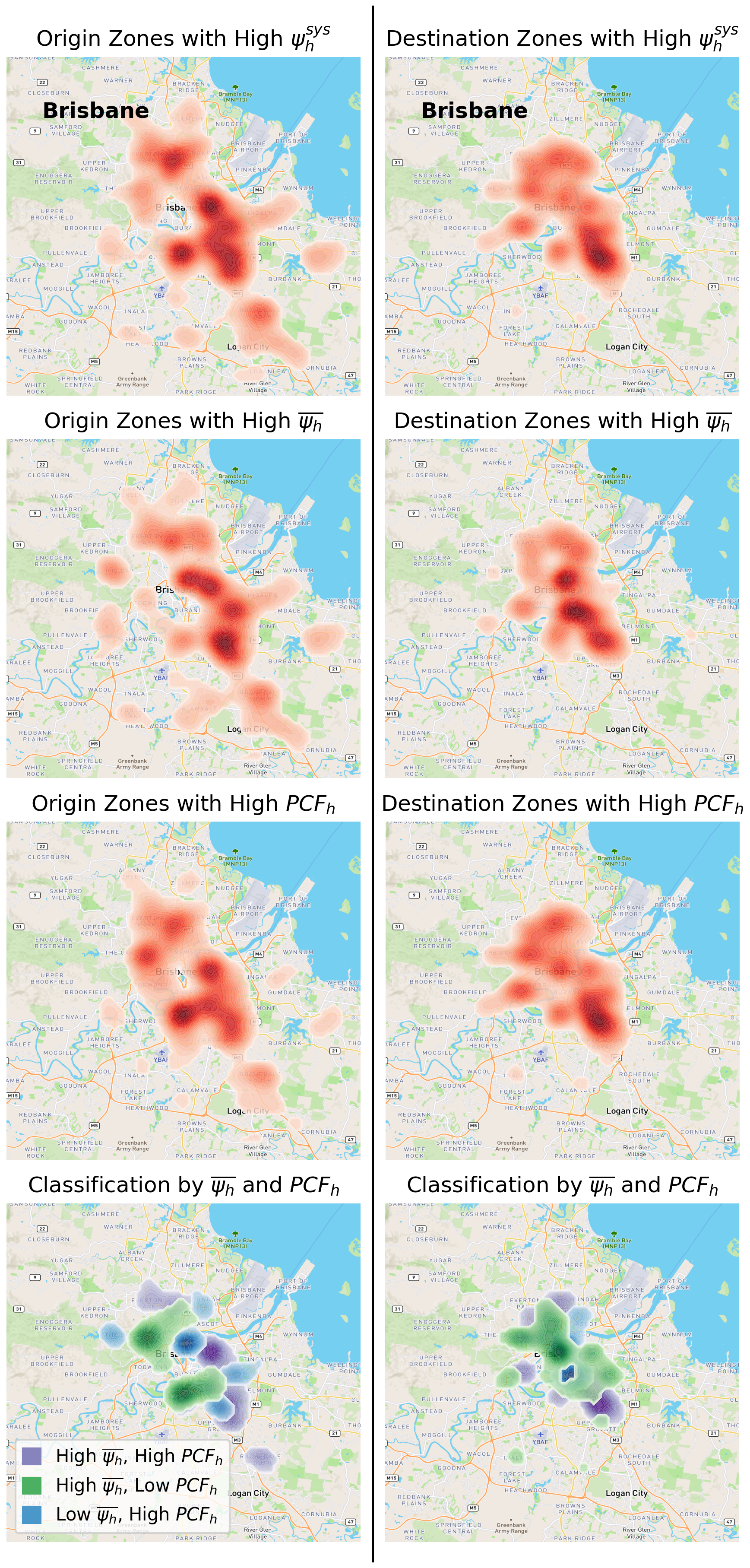}
    \caption{The spatial distribution of \textit{origin} zones with high \textit{departure time} peakedness and \textit{destination} zones with high \textit{arrival time} peakedness (AM period, Brisbane).}
    \label{fig:zone_contour_deparr}
\end{figure}

In addition to analysing origin zones with high departure time peakedness, as shown in Figure \ref{fig:zone_contour_city} in Section \ref{sec:results_impact}, we conducted a complementary analysis of destination zones by estimating the three peakedness measures $ \psi_h^{sys} $, $ \overline{\psi_h} $, and $ PCF_h $ from arrival time distributions. Using the same approach described in Section \ref{sec:results_impact}, we identified destination zones in Brisbane that exhibit high arrival time peakedness and compared them with the previously analysed origin zones. The spatial distribution of these destination zones is presented in Figure \ref{fig:zone_contour_deparr}. Similar to the origin-based analysis, the results reveal notable discrepancies between destination zones with high system-wide arrival time peakedness ($ \psi_h^{sys} $) and those with high user-level arrival time peakedness ($ \overline{\psi_h} $). Even though destination zones in Brisbane during the AM period are generally concentrated in the city centre, there are still considerable differences between destinations with high trip concentration based on aggregated arrival distributions (first row) and those where travellers individually exhibit a high degree of peakedness in their respective arrival time distributions (second row). This reinforces our earlier finding that system-wide demand patterns do not always align with individual users' travel time preferences. The distinct zone groups presented in the final row of Figure \ref{fig:zone_contour_deparr} further highlights the importance of considering both user-level peakedness ($ \overline{\psi_h} $) and peak window coincidence ($ PCF_h $) separately. By distinguishing between these measures, transit operators can better address the impact of destination disruptions in different scenarios, such as disruptions that impose high rescheduling or late arrival penalties on users with rigid arrival schedules (due to high $ \overline{\psi_h} $) and those that affect a large number of travellers simultaneously by disrupting a common peak period (due to high $ PCF_h $).

This analysis also provides additional insights into areas that exhibit high peakedness in both departure and arrival patterns by jointly analysing both origin and destination zones. These zones represent locations where rigid travel behaviour is observed both as origins and destinations. A disruption in these zones could impact a large number of people’s departures as well as a large number of people’s arrivals, amplifying the overall system-wide effect.

\end{document}